# Viroinformatics-based investigation of SARS-CoV-2 core proteins for potential therapeutic targets

***Lokesh Agrawal[1,2,3*], Thanasis Poullikkas[4,5], Scott Eisenhower[4,6], Carlo Monsanto[7], Ranjith Kumar Bakku[8,9]***

[1]*Universidad Integral del Caribe y América Latina, Kaminda Cas Grandi #79, Curaçao*
[2]*Graduate School of comprehensive human sciences, University of Tsukuba, 1-1-1 Tennodai, Tsukuba 305-8577, Japan*
[3]*Molecular Neuroscience Unit, Okinawa Institute of Science and Technology Graduate University, Kunigami-gun, Okinawa 904-0412, Japan*
[4]*Human Biology, School of Integrative and Global Majors, University of Tsukuba, 1-1-1 Tennodai, Tsukuba 305-0006, Japan*
[5]*Department of Experimental Pathology, Faculty of Medicine, University of Tsukuba, 2-1-1 Tennodai, Tsukuba 305-8576, Japan*
[6]*Department of Infection Biology, Faculty of Medicine, University of Tsukuba, 1-1-1 Tennodai, Tsukuba 305-8575, Japan*
[7]*Research Workgroup, Ronin Institute, 127 Haddon Place, Montclair, NJ 07043-2314 USA*
[8]*Faculty of Engineering Information and Systems, Dept. of Computer Science, University of Tsukuba, 1-1-1 Tennodai, Tsukuba, Ibaraki, 305-8572, Japan*
[9]*Tsukuba Life Science Innovation Program (TLSI), University of Tsukuba, 1-1-1 Tennodai, Tsukuba, Ibaraki, 305-8572, Japan*

***Corresponding Author**
*Email: lokesh.agrawal@unical.university*
*Mobile: +81-8094427502*
*ORCID Id: https://orcid.org/0000-0002-4165-1286*

## Abstract

Due to SARS-CoV-2 (Severe Acute Respiratory Syndrome Coronavirus 2) being a novel virus, there are currently no known effective antiviral drugs capable of slowing its progress. To accelerate the discovery of potential drug candidates, bioinformatics based *in silico* drug discovery can be applied as a very robust tool. In the present study, more than 60 antiviral drugs already available on the market, were chosen after literature survey. These can be used in clinical trials for the treatment of COVID-19. In this study, these candidate drugs were ranked based on their potential to interact with the Spike protein and RdRp (RNA-dependent RNA polymerase) of SARS-CoV-2. Additionally, the mechanism of their action as well as how the virus infection can utilize Hemoglobin to decrease the oxygen level in blood is explained. Moreover, multiple sequence alignments of the Spike protein with 75 sequences of different viruses from the Orthocoronavirinae subfamily were performed. This gives insight into the evolutionarily conserved domains that can be targeted using drug or antibody treatment. This multidimensional study opens a new layer of understanding about the most effective drug-targetable sites on the Spike protein of SARS-CoV-2.

## Introduction

In the past two decades, the world has faced several infectious disease outbreaks such as Influenza A (H1N1), SARS, MERS, Ebola, and Zika virus, which have had a considerable global impact on the healthcare delivery and economic systems. More recently, the outbreak of novel coronavirus / 2019-nCov / SARS-CoV-2, a newly discovered virus from the coronavirus family that causes COVID-19, has spread worldwide from the Chinese city of Wuhan (Cui, Li et al. 2019; Andersen, Rambaut et al. 2020; Shereen, Khan et al. 2020). SARS-CoV-2 is a highly transmittable coronavirus strain that can cause severe biological responses in among others the lower portion of the respiratory tract. Through the collection of epidemiological data, healthcare authorities have been able to determine the route through which person-to-person transmission of SARS-COV-19 has taken place, spreading at such high rates that it has become a worldwide public health challenge (Shereen, Khan et al. 2020). Many patients who tested positive for SARS-COV-19 who also had underlying comorbidity, have developed pneumonia-like symptoms. Especially elderly and patients with a compromised immune function, seem to be more susceptible to infection and subsequent mortality. In these high-risk patients, COVID-19 presents itself with flu-like symptoms which then rapidly worsen into severe pulmonary edema, respiratory distress blood thickening result stoke or cardiac arrest impose a serious threat to the life of COVID-19 patients (Avula, Nalleballe et al. 2020; Baldi, Sechi et al. 2020; Li and Ma 2020; Richardson, Hirsch et al. 2020) . Recently researchers also reported severe neurological problems in these patients which put a huge risk to the mental health in the society (Helms, Kremer et al. 2020; Mao, Jin et al. 2020). According to the latest report from the WHO (World Health Organization), up to the present, 27.7 million people were infected with COVID-19 and the death toll has reached 900 thousand worldwide. Of these, the American and European continents cover 6.3 million and 2.5 million cases respectively; together comprising around one third of known cases throughout the world. Despite cautious optimism related to the recent decline in cases and subsequent relaxation of containment procedures in various cities, there is still a fear of new waves of infections. Therefore, discovering an effective inhibitor of SARS-CoV-2 is urgently needed to protect human lives.

During the early stages of the global COVID-19 crisis, doctors scrambled to find an effective treatment for those suffering as a result of the coronavirus infection. Researchers are trying to develop and establish various treatment modalities such as plasma therapy, antibody treatment, decoy peptides/molecules and vaccine development (Duan, Liu et al. 2020; Inal 2020; Tobaiqy, Qashqary et al. 2020; Zeng, Chen et al. 2020). Unfortunately, the exact mechanisms of its pathogenesis remain unclear, which has made finding an effective treatment very difficult. To find a cure for COVID-19 we must first understand its structure and infection mechanism. SARS-CoV-2 is a contagious virus that can be spread through inhalation or ingestion of moisture droplets that contain the virus, resulting from coughing and sneezing (Zhang, Li et al. 2020). Structurally, SARS-CoV-2 is a single-stranded RNA virus with a genome that consists of ∼30000 nucleotides. This encodes four structural proteins: Nucleocapsid (N) protein, Membrane (M) protein, Spike (S) protein, Envelop (E) protein, and ∼16 non-structural proteins (nsp) (Schoeman and Fielding 2019; Boopathi, Poma et al. 2020; Jin, Du et al. 2020). The viral surface proteins, Spike, envelope, and membrane are embedded in a lipid bilayer enveloping a capsid protein shell in which there is a nucleocapsid protein (N-protein) that is bound to the virus's single positive-strand RNA genome (Boopathi, Poma et al. 2020). The M-protein is most abundant in the viral surface and it is believed to be the central organizer for the coronavirus assembly. The trimeric form of Spike protein is integrated across the surface of the virus; it mediates attachment of the virus to the host cell surface receptors and fusion between the viral and host cell membranes to facilitate entry into the host cell (Vinson 2020; Wrapp, Wang et al. 2020). The coronavirus Spike protein is subjected to proteolytic cleavage by host proteases (i.e. trypsin and furin) in two sites at PCS (Polybasic furin Cleavage Site) located at the boundary between the S1 and S2 subunits (S1/S2 site) (Fig. S1 and 9). Binding of ACE2 with Spike S1 (AA: 13-685) proteins allows the virus to adhere to lung epithelial cells and through a tube composed of Spike S2 proteins (AA:686-1273) it injects the genetic material inside the host cell (Fig. S2 and S3). Once the virus reaches the budding stage, the S2 domain (S2′ site, AA: 816-1273) is cleaved to release the fusion peptide (Shang, Ye et al. 2020). The E-protein is a small membrane protein composed of ∼76 to 109 amino-acid and minor components of the virus particle. It plays an important role in virus assembly, membrane permeability of the host cell, and virus-host cell interaction. A lipid envelope encapsulates the genetic material. In addition to these, Hemagglutinin-esterase dimers (HE) have been located at the surface of the viral capsid (Zeng, Langereis et al. 2008). Although HE proteins may be involved in virus entry, they are not required for replication; however, they do appear to be important for infecting natural host cells.

With the current need to reduce the cost, time, and risks of the drug development process, scientists are exploring the use of already approved drug candidates to test in COVID-19 patients such as Hydroxychloroquine, Remdesivir, and Azithromycin (Gautret, Lagier et al. 2020; Grein, Ohmagari et al. 2020; Rosenberg, Dufort et al. 2020). The ability to reuse FDA approved/preclinical trial drugs to target multiple virus proteins of SARS-CoV-2 provides a ray of hope for the survival of humanity. In this work, we thoroughly investigated the potential use of broadly available antiviral, anti-bacterial, anti-parasitic drugs, flavonoids, and vitamins, which have shown antiviral activity. After thorough literature survey we selected these drugs which can be used for the treatment of COVID-19, here we explore their interaction efficiencies with Spike and RDRP proteins of the virus. Computational simulation is an essential tool to accelerate the discovery of therapeutics which can effectively cope with the rapidly evolving structure of SARS-CoV-2 on the molecular level (Boopathi, Poma et al. 2020). In addition, clinical trial reports suggest that patients suffering from COVID-19 have a low oxygen concentration in their blood not caused solely by respiratory distress (Kashani 2020; Wang, Hu et al. 2020; Xie, Covassin et al. 2020). This phenomenon led us to the hypothesis of a possible binding event between SARS-CoV-2's Spike protein and the hemoglobin, specifically heme group interaction (Cavezzi, Troiani et al. 2020; liu and Li 2020). Further, to find out the therapeutically important conservative protein motifs of the Spike protein, we performed multiple sequence alignments between the Spike protein sequences of SARS-CoV-2 and 75 sequences from different virus species that belong to the Orthocoronavirinae subfamily.

## Materials and Methods

### 1) Docking of antivirals, antibiotics, antiparasitics, flavonoids and vitamins with Spike and RdRp protein of SARS-CoV-2

3D models of SARS-CoV-2 Spike (QHD43416.pdb) and RdRp (QHD43415_11.pdb) proteins were downloaded from Zhang Lab I-TESSAR online library in .pdb format (Yang and Zhang 2015). 3D structures of drugs and small molecules for docking were downloaded from PubChem in .pdb format. There was no change observed in the structure of these drug molecules after getting activated inside the liver, which makes them a suitable candidate for the docking study. Docking was performed with AutoDockTool (v1.5.6) software (Morris, Huey et al. 2009). All models were converted to. pdbqt format after deletion of water molecules and computing Gasteiger algorithm for macromolecules, whereas in ligands after choosing torsions $\leq 9$. For developing the grid box for SARS-CoV-2 Spike protein we set the following metrics: X-dimension: 104, Y-dimension: 116, Z-dimension: 126, Spacing: 1.000, Offset X: 2.917, Offset Y: 1.450 and Offset Z: 0.194 and grid were set using Autogrid4. Docking was performed with Autodock4, for each docking we set our macromolecules as a fixed filename and docked with Genetic algorithm and output set to Lamarckian GA (4.2). After the docking simulation, 10 docking models were analyzed and ranked by their energy binding ability ($\Delta G$). H-bonds were also built in the docking models and the best model of each docking was saved as .pdbqt format file.

Docking models were further analyzed for the discovery of a binding pocket in Discovery Studio Visualizer (DSV) (v2019 client; Dassault Systèmes BIOVIA) software. Interactions were built by selecting all favorable interaction types intermolecularly. Furthermore, 2D diagrams were visualized through DSV.

### 2) *In silico* protein-protein interaction

Protein-protein docking was performed through the ClusPro (v2.0) server (Kozakov, Hall et al. 2017). Macromolecules were submitted in .pdb format and the most favorable model was downloaded for further analysis in DSV. We docked the SARS-CoV-2 Spike protein with both the tetramer and monomer molecules of Hemoglobin (1A3N.pdb) and the Heme group (extracted from 1A3N.pdb), in ClusPro server and AutoDockTools (v1.5.6) software respectively. LigPlot + v.2.2 software was used to generate a 2D plot of interface amino acids at the Hemoglobin-Spike interaction pocket (Laskowski and Swindells 2011).

### 3) Finding of most interacting motifs of Spike protein with antiviral drugs and heat-map representation of motifs

We manually investigated the binding pocket of the Spike protein and RdRp with the docked drug ligands. We divided the whole protein sequence of Spike and RdRp proteins in 12 and 10 equally spanned motifs respectively. Based on the number of interactions with the docked drug ligand database, we ranked each motif and further normalized the number of interactions in each motif to compare their potential as a target for the therapeutic drugs and their significance for the future drug development against SARS-CoV-2. In continuation, based on the normalization of the total number of interactions for each motif, heat-maps were produced using DSV. Similarly, we investigated the frequency and location of specific amino acids of Spike and RdRp in each motif, which is important for the interaction with various drug ligands.

### 4) Investigation of conservative motifs of SARS-CoV-2 Spike protein and phylogenetic analysis

To fulfill this purpose, we used the Blast Molecular Evolutionary Genetics Analysis (MEGA 10.0.5) tool. In the search of unique conservative regions in SARS-CoV-2 Spike protein throughout its evolution, we performed alignment of the amino acid sequence of the Spike protein with BLASTP database and selected 75 subject sequences different viruses from the Orthocoronavirinae subfamily (Fig.S5), which have the highest BLASTP score. Later, we ran the sequence alignment of these 75 subject sequences from the database using the MEGA-X software's MUSCLE program (Kumar, Stecher et al. 2018). Further, a phylogenetic tree was generated based on the Fast-Minimum Evolution tree clustering method and the Grishin distance method of MEGA-X. The phylogeny was tested using the resampling bootstrap methodology with 1000.

## Results

### 1. Potential drugs for the treatment of COVID-19

### 1.1. Most potent antiviral drug targeting of SARS-CoV-2 Spike protein

In our study, we focused on 47 different antiviral agents (Fig. 1A) most readily available on the market. AutoDock based docking results suggested their interaction efficiency/binding efficiency with Spike protein of SARS-CoV-2 along with binding pocket and location of most important amino acids which are crucial for drug interaction. We have shown the top heat antiviral drugs based on their efficiency of interaction (Fig.2; Table S1) against the Spike protein; starting from the highest binding efficiency : Indinavir ($\Delta G$ = -9.8 kcal/Mol), Nelfinavir ($\Delta G$ = -9 kcal/Mol), Fosamprenavir ($\Delta G$ = -8.2 kcal/Mol), Rintatolimod ($\Delta G$ = -7.6 kcal/Mol), Loviride ($\Delta G$ = -7.3 kcal/Mol), Nevirapine ($\Delta G$ = -7.3 kcal/Mol), Nitazoxanide ($\Delta G$ = -7.1 kcal/Mol), Imiquimod ($\Delta G$ = -6.8 kcal/Mol), Inosine ($\Delta G$ = -6.5 kcal/Mol), and Cobicistat ($\Delta G$ = -6.37 kcal/Mol). Although, most of the antivirals bind with the S1-NTD region and its surroundings, which suggest they might physically block the trimeric assembly of the Spike protein and/or the viral adhesion to the host cell membrane via blocking the interaction between S1 and cell surface receptor such as ACE2.

### 1.2. Most potent antibacterial/antiparasitic/flavonoids/vitamins for the targeting of SARS-CoV-2 Spike protein

In our study, we focused on 21 different antibiotics, 3 antiparasitic (including hydroxychloroquine), 5 flavonoids, and 3 small ligands including Vitamin C and Vitamin D (Fig.1B); Which were proven to have some effect on other coronaviruses, and they are broadly accessible. Analyses of the AutoDock results yielded a list of the top antibiotics based on their efficiency of interaction (Fig.3; Table S2) with the Spike protein that is as follows : Vancomycin (ΔG = -10.2 kcal/Mol), Nelfinavir (ΔG = -8.6 kcal/Mol), Azithromycin (ΔG = -7.9 kcal/Mol), Sulfamethoxazole (ΔG = -6.6 kcal/Mol), Meropenem (ΔG = -6.4 kcal/Mol), Tenofovir Disoproxil (ΔG = -6.3 kcal/Mol), Trimethoprim (ΔG = -6.1 kcal/Mol), Ciprofloxacin (ΔG = -5.58 kcal/Mol), Gentamicin (ΔG = -5.4 kcal/Mol), and Levofloxacin (ΔG = -5.11 kcal/Mol).

Similarly, we found the top antiparasitic drugs (Fig. 1B) that interact with the Spike of SARS-CoV-2 (Fig.4A-4B; Table S3), which are as follows: Ivermectin B1a (ΔG = -9.16 kcal/Mol), Ivermectin B1b (ΔG = -8.86 kcal/Mol), and Hydroxychloroquine (ΔG = -2.85 kcal/Mol). Furthermore, in our study, we focused on 5 different flavonoids (Fig. 1B) that were proven to have some effect on other coronaviruses. AutoDock results of the candidates portrays (Fig.4C-4D; Table S3) Tetramethoxyflavone as the peak (ΔG = -4.91 kcal/Mol), with Herbacetin (ΔG = -4.74 kcal/Mol) and Gallocatechin (ΔG = -2.88 kcal/Mol) following afterward. Finally, we investigated the efficiency of vitamins as a therapeutic for the treatment of COVID-19 (Fig. 4E-4F; Table S3). We found that Vitamin D (ΔG = -5.52 kcal/Mol) and Vitamin C (ΔG = -2.95 kcal/Mol) can interact with the Spike protein.

### 1.3. Most potent antiviral drug targeting of SARS-CoV-2 RdRp protein

Our investigation included an examination of multiple drugs targeting viral RdRp and then comparing that to the activity of those same drugs on the Spike protein of SARS-CoV-2 (Table S4). We found that the top antiviral drugs having interaction with RdRp are as follows: Beclabuvir (ΔG = -5.63 kcal/Mol), Galidesivir (ΔG = -4.38 kcal/Mol), Ribavirin (ΔG = -4.2 kcal/Mol), Favipiravir (ΔG = -3.09 kcal/Mol), Sofosbuvir (ΔG = -2.95 kcal/Mol), Tenofovir (ΔG = -2.34 kcal/Mol), and Remdesivir (ΔG = -1.28 kcal/Mol). Most interestingly, the binding potential of these drugs for the interaction with Spike seems to be very similar to interaction with RdRp in terms of the formation of the most energetically favorable complex (Fig. 5A). As the subject that sparked interest in the research, Favipiravir was observed to bind to the RdRp molecule with an energy of -3.09 kcal in the amino acid region ranging from 380 to 510. Additionally, when examined against the Spike protein monomer, Favipiravir was observed to bind to the protein within the 620 to 650 amino acid range at a bond energy of -2.94 kcal. Using Favipiravir as a standard, the six most prominent RdRp-targeting drugs on the market were likewise examined for their binding properties. Of the drugs examined, most stayed within a binding energy range of ±2.0 kcal of Favipiravir, but notably, the drug Beclabuvir significantly exceeded this with an energy of -5.63 kcal to the amino acid range 130 to 190 when binding to RdRp and -6.64 kcal to the amino acid ranges 220 to 300 and 600 to 640 when binding to the Spike monomer.

## 2. Most important motifs of RdRp protein as a target for therapeutic drugs

Most of the drugs targeting RdRp mainly interact between the amino acid residues 101-200 and 701-800 specially THR-51, ASN-122, PHE-133, LYS-182, CYS-760, THR-761, ALA-771, GLU-780 (Fig.5B). Additionally, we also found the subtle interaction between 1-100 and 801-900 amino acid residues.

## 3. Most important motifs of Spike protein of SARS-CoV-2 as a target for therapeutic drugs

Most of the antiviral drugs interact with Spike S1-NTD especially between the residues 1-100 and 201-300 (Fig.6A). Most antiviral drug interacting residue locations at S1 are VAL-3, PHE-4, VAL-6, PHE-58, and PRO-82. On the other hand, most of the antiviral interaction occurs with Spike S2 lies between the residues 801-1000, and 1201-1300. Especially the amino acids ASN-824, VAL-826, THR-827, LEU-945, LYS-1205, TYR-1209, PRO-1213, and TRP-1217 are very important. In continuation, interface residues where most of the antibiotics interact also lie on the S1-NTD between 1-100 and 201-300 amino acids which is similar to the antivirals (Fig.6B). Specifically, the residues VAL-3, VAL-6, LEU-7, LEU-8, ARG-21, GLN-23, LEU-24, PRO-26, PRO-82, VAL-83, and VAL-289. We also found the interaction pocket at S2 especially residues between 701-800, 901-1000, and 1201-1300 such as GLY-744, ASP-745, LEU-966, VAL-976, LEU-977, ASN-978, and ARG-1000 are very important in the interaction with antibiotics. In addition, residues 601-700 at the junction of S1 and S2 seem important for the action of antibiotics.

Furthermore, in the case of antiparasitic drugs, we found that the residues of S2 Spike between 901-1000 and 1101-1200 are very important especially the locations LYS-1205 and TYR-1209 are similar to antivirals and antibiotics (Fig.6C). However, we also found interaction at S1-NTD between the residues 1-101 and 200 which suggests the therapeutic importance related to S1 Spike. Additionally, a study with flavonoids revealed the interaction at S1-NTD between the residues 1-100 especially the locations VAL-6 and ARG-21, and at S2 between the residues 701-900 such as LYS-790, LYS-811, ALA-893 and LEU-1224 (Fig.6D). Similarly, the interaction with Vitamins C and D lies between the residues 1-100, 301-400, and 501-600 amino acids of Spike S1, most importantly VAL-47, PHE-318, and THR-630, and between the 701-1000 residues of Spike S2 especially residues TYR-741, ILE-742, CYS-743, and GLY-744 (Fig.6E).

## 4. Interaction between Hemoglobin and Heme protein with Spike of SARS-CoV-2

We docked the SARS-CoV-2 Spike protein with both the molecule of hemoglobin and the heme group in the ClusPro (v2.0) server and AutoDockTools (v1.5.6) software, respectively. Taking into consideration the docking simulation of hemoglobin, both the monomer (Fig.7) and the tetramer (Fig. S4) of it bind very tightly to S1-NTD with several hydrogen bonds bridging the two molecules. Residues of the Spike protein forming hydrogen bonds with the hemoglobin monomer are ARG-237, ASN-37, LEU-270, ARG-88, ASP-627, HIS-625, ASN-61, THR-63, GLN-271, GLN-23, and TYR-28 (Fig.7). Similarly, we found that ARG-237, ARG-21, VAL-53, VAL-6, PHE-4, VAL-3, ASP-111, SER12, GLN-14, ARG-158, CYS-15, THR-22, THR-73, THR-74, GLU-132, LYS-113, and SER-112 of the Spike protein form a hydrogen bond with the tetramer of hemoglobin (Fig. S4). Interestingly, the ARG-237 hydrogen bond persists in both simulations (monomer and tetramer), thus indicating a possible binding area that is more favorable for the interaction in terms of binding efficiency and spatial pairing. Moreover, the region surrounding the above-mentioned amino acid might facilitate the binding of the heme group to the S1-NTD region of the Spike protein. Simulation of this interaction was achieved through AutoDock software and yielded a binding energy score of -5.4 kcal/Mol. Results of interaction show several Vander Waals bonds surrounding the heme group, whereas 6 Pi-(Anion, Sigma, -Pi Stacked and Alkyl) bonds and 1 hydrogen bond bridging with Spike protein (Fig.8). Amino acid residues TYR-38, LYS-41, PHE-43, LYS-206, GLU-224, PRO-225, LEU-226, VAL-227, ASP-228, and THR-284 of Spike protein interact with the heme group directly. The core amino acid of this interaction is GLU-224 that might possess the ability to bind directly to the heme iron with a Pi-Anion bond and thus preventing the free oxygen-binding into the area.

## 5. Conservative motifs of SARS-CoV-2 Spike protein and phylogenetic analysis correlation between the sequence of Spike protein

In the search of finding unique regions in the SARS-CoV-2 Spike protein, we performed an alignment of the amino acid sequence of the Spike protein with the BLASTP database (Fig.S5). Top 5 sequences that showed high alignment scores were aligned again using MEGA-X software and a detailed analysis of five important Spike protein domains were examined. Results suggest that conservation exists in higher levels at S2-HR1 and HR2, moderate conservation at S1-NTD, and PCS, whereas at S1-CTD there are many hypervariable regions (Fig. 9). We also ran the sequence alignment of 75 subject sequences from the database using MEGA-X software and developed a phylogenetic tree through BLASTP. The subject sequences suggested by BLASTP are related to Spike proteins from MERS, Murine hepatitis virus (MCoV), Human coronavirus (HCoV), Feline Infectious Peritonitis virus (FIP-CoV) and SARS-CoV-1 (Fig. 10). Initial results suggest that the SARS-CoV-2 Spike protein shares similarities between SARS-CoV-1; whereas a common ancestor connects MERS and SARS-CoV-2. The majority of similar sequences aligned with SARS-CoV-2 Spike protein are from the Betacoronavirus of the subspecies of Sarbecoviruses, Merbecoviruses, and Embecoviruses; conversely, the minority of them come from Alphacoronaviruses.

For the alignment of the 75 subjected sequences, we could identify several conserved regions spanning throughout the protein. Several "clusters" of conserved regions spanning from the beginning S1-CTD until its end at the 330 to 529 amino acid range. That means S1-CTD can be divided into 7 clusters of conserved domains, composed of around 25 amino acids each, according to the MEGA-X software alignment. This gives a deeper insight into the nature of the Spike protein Binding Domain to ACE2 and other receptors. Moreover, two massive evolutionarily conserved domains lie at the center of the S2 region within the amino acid ranges of 908 to 1003 and 1163 to 1211; indicating that the S2 part of Spike protein seems to be more heavily conserved compared to its S1 counterpart. In conclusion, the S1-CTD and the final portion of the S2 protein are relatively conserved throughout the species and they will be effective references for the discovery of a universal drug against all strains of SARS-CoV-2.

## Discussion

### 1. Importance and Significance of drug docking with Spike

#### 1.1. Antivirals

At the onset of the pandemic, doctors began to try a plethora of antiviral drugs with the likelihood of success in an attempt to slow the progression of COVID-19 in those affected by the infection. However, it would be both illogical and immoral to try all of these drugs on human patients without any knowledge as to whether or not they will have any interaction with the SARS-CoV-2 virus. In our study, we focused on 47 different antivirals already approved and broadly available on the market that are thought to have some potential for inhibiting the virus. Seeing as the Spike protein is evolutionarily conserved, it is a prime location to observe for interaction and inhibition of the virus in a manner that is ubiquitous across all potential strains. Using the AutoDock software, we performed docking simulations between these drugs to observe the energy/efficiency of their interactions with the SARS-CoV-2 Spike protein. Arranged by their binding energy (Fig. 1A), these drugs show a wide range of effectivity from Saquinavir at the lowest ($\Delta G$ = -0.23 kcal/Mol) to Indinavir at the highest ($\Delta G$ = -9.8 kcal/Mol). Drugs with interaction energy greater than or equal to the Japanese-made Favipiravir ($\Delta G$ = -5.2 kcal/Mol) are prime candidates to be potential treatments as it has already been observed to have some mild effects on inhibition of COVID-19 in patients (Wang, Cao et al. 2020).

Further examination of the AutoDock results revealed that the drug molecules interacted with the Spike protein monomer at very specific locations (Fig. 6A). Most notable are the β-sheets between 1-100 residues range having over 150 different molecular interactions. This is followed by the α-helix between 801-900 residues range yielding over 100 unique interactions. Noting the location of the 1-100 range, drugs that interact with that binding pocket will most likely not have a large-scale effect on the protein's functionality as it only serves as the N-terminal domain. Conversely, the 801-900 region is located at the point on the Spike monomer that is key to the formation of the Spike trimer; binding to this region would impair SARS-CoV-2's ability to form the trimeric structure and overall inhibit the production of viable virions (Vinson 2020; Wrapp, Wang et al. 2020).

#### 1.2. Antibiotics

Normally, when looking at stopping a virus, antibiotics would not be considered due to their specific targeting of bacterial processes. Overuse of antibiotic treatments on non-bacterial agents has been a major factor that has led to the rise of drug resistance (Murray 2020). However, all options must be considered in an emergency such as the COVID-19 pandemic and bioinformatics tools have the luxury to examine the potential of antibiotics being used against SARS-CoV-2 without the possibility of contributing to drug resistance. Additionally, researchers have shown the combinational therapy of Azithromycin and Hydroxychloroquine suppressed the growth of SARS-Cov-2 virus *in-vitro* (Gautret, Lagier et al. 2020). Therefore, similar to our examination of antiviral drugs, we investigated the binding properties of 21 prominent antibiotics to the SARS-CoV-2 Spike protein using the AutoDock software. After arranging the drugs based on their binding energy (Fig.1B), a steady increase in energy can be observed from Rifampicin at the lowest (ΔG = -1.83 kcal/Mol) to Vancomycin at the highest (ΔG = -10.2 kcal/Mol). As was previously stated, drugs with binding energy higher than ΔG = -5.2 kcal/Mol exhibit inhibitory properties on the virus (as seen with Favipiravir), therefore 9 antibiotics beginning from Gentamicin (ΔG = -5.4 kcal/Mol) have potential to disrupt viral function. Unfortunately, the majority of the molecular interactions observed occur in the N-terminal domain of the Spike protein (Fig.6B) which indicates that they might yield little to no result in disrupting the virus.

### 1.3. Antiparasitic

Researchers have reported that Ivermectin and Hydroxychloroquine, which are FDA-approved anti-parasitic drugs previously shown to have broad-spectrum antiviral activity *in-vitro* (Gautret, Lagier et al. 2020; Heidary and Gharebaghi 2020). Leon Caley et al. showed that with a single addition of Ivermectin to Vero-hSLAM cells 2 h post-infection with SARS-CoV-2 able to effect ~5000-fold reduction in viral RNA at 48 h (Caly, Druce et al. 2020). Ivermectin, therefore, warrants further investigation for possible benefits in humans. Similarly, the 4-aminoquinoline antimalarials chloroquine and Hydroxychloroquine have been promoted and sometimes used in the treatment of COVID-19, alone or combined with Azithromycin, based on their immunomodulatory and antiviral properties, despite an absence of methodologically appropriate proof of their efficacy (Gautret, Lagier et al. 2020).

In our study, we examined the potential for these antiparasitic drugs to interact with the SARS-CoV-2 Spike protein using the AutoDock software. By comparing the binding energies that resulted from this analysis (Fig.1B) it can be observed that Hydroxychloroquine has a relatively low energy interaction (ΔG = -2.85 kcal/Mol) while Ivermectin B1a and B1b have high energy bonds (ΔG = -9.16 kcal/Mol and ΔG = -8.86 kcal/Mol respectively). Additionally, these interactions are observed to occur primarily on the ɑ-helices existing in the 901-1200 amino acid region of the Spike monomer (Fig. 6C). Considering that this region is involved in the formation of the Spike trimer, specifically the base of Spike where protein is integrated into the virion and are conserved among coronavirus species (Shereen, Khan et al. 2020; Vinson 2020; Wrapp, Wang et al. 2020).

### 1.4. Flavonoids

Neglected from the wide scientific view, flavonoids played a crucial role many times against emerging viruses. Tetramethoxyflavone from elderberry extract inhibited Human Influenza A (H1N1) infection in vitro with an IC50 value of 252 ± 34 µg/mL (Roschek, Fink et al. 2009). Moreover, flavonoids were utilized against the infectious bronchitis virus (IBV), a pathogenic chicken coronavirus (Chen, Zuckerman et al. 2014). Whereas there is a non-well effective vaccine yet; Chen and colleagues tested non-cytotoxic, crude ethanol extracts of Sambucus nigra fruit for anti-IBV activity because it contains polyphenol derivatives that inhibit other viruses. Electron microscopy of virions treated with S. nigra extract showed compromised envelopes and the presence of membrane vesicles, which suggested a mechanism of action. These results demonstrate that S. nigra extract can inhibit IBV at an early point in infection, probably by rendering the virus non-infectious. They also suggest that future studies using S. nigra extract to treat or prevent IBV or other coronaviruses are warranted (Chen, Zheng et al. 2014). Additionally, coronaviruses of SARS and MERS have been rising targets of some flavonoids. The antiviral activity of some flavonoids against CoVs is presumed directly caused by inhibiting 3C-like protease (3CLpro). It has been suggested that flavonoids such as herbacetin, rhoifolin, and pectolinarin were found to efficiently block the enzymatic activity of SARS-CoV 3CLpro (Jitendra Subhash, Aroni et al. 2020; Jo, Kim et al. 2020).

In our study, we focused on 5 different flavonoids (Fig.1) that were proven to have some effect with other coronaviruses. AutoDock results of the candidates suggested a top hit of Tetramethoxyflavone ($\Delta G$ = -4.91 kcal/Mol) and a second top hit being Herbacetin ($\Delta G$ = -4.74 kcal/Mol). Although they did not exhibit high $\Delta G$ values, flavonoids showed unique binding areas. From the binding areas, we could roughly identify the preferent binding regions on the Spike protein (Fig.5). Total interactions per segment results suggest a tendency of flavonoids to bind in both S1-NTD and S2. Even though most of the screened drugs and small molecules bind with the S1-NTD region, flavonoids showed an increased number of interactions in the conserved S2 region too. This suggests that they might possess the ability not only for structurally inhibiting the assembly of the Spike trimer but also for binding into the lumen of the trimeric protein and thus inhibiting the viral fusion with the host cell in a variety of coronaviruses.

### 1.5. Vitamins

Earlier, researchers claimed that vitamin C has a beneficial effect on asthma. In the absence of a specific treatment for SARS, the possibility that vitamin C may show non-specific effects on severe viral respiratory tract infections should be considered (Hemilä 2003). Numerous reports are indicating that vitamin C may affect the immune system for example the function of phagocytes, transformation of T lymphocytes, and production of interferon (Leibovitz and Siegel 1981; Hemilä and Douglas 1999). In particular, vitamin C increased the resistance of chick embryo tracheal organ cultures to infection caused by an avian coronavirus (Atherton, Kratzing et al. 1978). Studies in animals found that vitamin C modifies susceptibility to various bacterial and viral infections, for example protecting broiler chicks against an avian coronavirus. Five placebo-controlled trials have shown quite consistently that the duration and severity of common cold episodes are reduced in the vitamin C groups, indicating that viral respiratory infections in humans are affected by vitamin C levels. There is also evidence indicating that vitamin C may affect pneumonia (Hemilä 1997). In particular, three controlled trials with human subjects reported a significantly lower incidence of pneumonia in vitamin C-supplemented groups, suggesting that vitamin C may affect susceptibility to lower respiratory tract infections under certain conditions.

Similarly, the role of Vitamin D has been proved very important for the prevention of lower respiratory tract diseases caused by viral infections. There are two forms of vitamin D, cholecalciferol (vitamin D3), and ergocalciferol (vitamin D2) (Yang, Lv et al. 2019). Vitamin

D has also been implicated as a causal factor for the common cold (Rondanelli, Miccono et al. 2018). Two observations point in this direction. Colds are more frequent in the winter when vitamin D levels are generally lower, and patients with chronic obstructive pulmonary disease who have lower levels of vitamin D have more upper respiratory tract infections (URIs). Citing the role of vitamin D, Gindi et al found that the number of recent URIs was inversely related to serum 25(OH)D levels; this linkage was stronger in patients with pre-existing respiratory tract disease (Ginde, Mansbach et al. 2009). On a more positive note, the Cochrane Collaborators' review of vitamin D for asthma is quite encouraging (Martineau, Cates et al. 2016). Recently, researchers found an association between mean levels of vitamin D in various countries and cases respectively mortality caused by COVID-19. Their results suggested that the mean level of vitamin D (average 56mmol/L, STDEV 10.61) in each country was strongly associated with the number of cases/1M (mean 295.95, STDEV 298.73 $p=0.004$, respectively with the mortality/1M (mean 5.96, STDEV 15.13, $p < 0.00001$) (Ilie, Stefanescu et al. 2020). Vitamin D levels are severely low in the aging population, especially in Spain, Italy, and Switzerland. This is also the most vulnerable group of the population for COVID-19. In addition to that, in the present study, we also found that vitamin C ($\Delta G$ = -2.95 kcal/Mol) and vitamin D3 ($\Delta G$ = -5.52 kcal/Mol) can directly interact with Spike protein of SARS-CoV-2 (Fig. 4E-4F). This finding lends the idea that vitamin C and vitamin D both can block the binding of Spike with host cell receptors, the formation of its trimeric form and fusion to host cells make them important candidates for further clinical trials to probe their potential for the treatment of lower respiratory tract diseases.

## 2. Importance and Significance of drug docking with RdRp

During the early stages of the SARS-CoV-2 pandemic, doctors scrambled to find an effective treatment for those suffering from the virus. Among the many drugs used in an attempt to slow the virus, the Japanese anti-influenza drug Favipiravir (sold as Avigan) showed promise when applied to COVID-19 patients. This is believed to be through its inhibition of the RNA-dependent RNA polymerase (RdRp) employed by RNA viruses during their replication process (Furuta, Komeno et al. 2017; Cai, Yang et al. 2020). However, as the pandemic wore on it became clear that this drug was only effective when administered at early stages of coronavirus infection; a phenomenon observed in tests against other viruses including alphaviruses, arenaviruses, and flaviviruses (Furuta, Komeno et al. 2017). This suggests that the inhibition of RdRp works primarily as a preventative measure rather than a primary form of treatment against RNA-based viruses. Further, our findings suggest the efficacy of Beclabuvir, in particular, to be used against SARS-CoV-2. Not only does it interact significantly with the viral RdRp, but it's binding energy to the Spike protein exceeds that of a significant number of the drugs tested in this study. This suggests that it would have a stronger inhibitory factor to the virus than that of Favipiravir and provide additional protection in the form of disrupting the Spike protein's function.

In our investigation, we examined 7 antiviral drugs known to primarily interact with RdRp in their respective viral targets. Using the AutoDock software to generate docking simulations of these drugs to the SARS-CoV-2 RdRp molecule, the drugs were arranged according to their binding energies (Fig. 5A). Using Favipiravir as a standard at $\Delta G$ = -3.09 kcal/Mol for the ability to exhibit a mild clinical success, 3 RdRp-targeting drugs are immediately identifiable as potentially more successful: Ribavirin (ΔG = -4.20 kcal/Mol), Galidesivir (ΔG = -4.38 kcal/Mol), and Beclabuvir (ΔG = -5.63 kcal/Mol). Additional examination of where these molecular interactions take place revealed that they are primarily concentrated in the 101-200 and 701-800 amino acid regions (Fig. 5B). Similar to the Spike protein, the 701-800 regions in particular consist of a series of α-helices that are essential in the RdRp's functionality; blocking them would, therefore, inhibit the use of RdRp and essentially shut down the virus's ability to replicate its genome.

## 3. Significance of Heme/hemoglobin and Spike interactions

Recently many anecdotal clinical trial reports suggest that patients affected by COVID-19 not only have impaired respiratory function but also low oxygen count causing respiratory distress (Li and Ma 2020; Wang, Hu et al. 2020; Xie, Covassin et al. 2020). This phenomenon led us to the hypothesis of a possible binding event between SARS-CoV-2 Spike protein and the heme group of the hemoglobin molecule. That binding might greatly reduce the oxygen affinity to hemoglobin and increase its degradation rate as many COVID-19 survivors showed much less hemoglobin count compared with the non-infected (Cavezzi, Troiani et al. 2020; liu and Li 2020). It was also revealed that alternative treatments currently being considered for COVID-19 such as Chloroquine and Hydroxychloroquine by increasing hemoglobin production and increasing hemoglobin availability for oxygen binding and acetazolamide by causing hyperventilation with associated increasing levels of oxygen and decreasing levels of carbon dioxide in the blood may significantly ameliorate COVID-19 respiratory symptoms (Iyamu, Perdew et al. 2009; Pastick, Okafor et al. 2020).

Considering the docking simulation of hemoglobin, Spike protein has the possible potential to interact with the tetramer and monomer of it (Fig.7). Amino acid GLU224 might be the key target as it plays a crucial role of binding directly to the iron particle of heme group. This is the first time shown that the Spike protein of SARS-CoV-2 might interact with not only the hemoglobin molecule but as well as with the heme group directly to its iron particle. These results will shed light on the, yet inkown, intrinsic mechanism of SARS-CoV-2 Spike protein and the low oxygen count of patients and survivors, whereas at the same time offers a possible target side for the development of new drugs.

## 4. Alignment and drug comparison:

We also identified conserved regions through the sequence alignment and found several portions of the protein that have some conservation among similar species Spike proteins. Starting from S1-NTD, 10 out of 75 subjected sequences match with its first portion at 1-200 amino acid range, whereas 24 out of 75 sequences match with the S1-NTD's last portion at 201-300. A moderate amount of similarity was observed at the S1-CTD side with 33 similar sequences at around 300-510. Moreover, the highest conservation among subjected sequences was observed in the center of the S2 portion of the protein at around 900-1000 in a number of 42 from 75 (Fig. S5). To strengthen these results, we aligned 5 Spike proteins among several species related to coronavirus (Porcine epidemic virus, MERS-CoV, FIP-CoV, SARS-CoV-1, and Human-CoV) and investigate their conservation in five core domains (S1-NTD, S1-CTD, PCS, S2-HR1, and S2-HR2). Domains of S1-NTD, HR1, and HR2 showed high conservation (Fig. 9). This might suggest that drugs which tend to bind to those regions are more effective in addressing a bigger spectrum of coronaviruses. Interestingly, most of the antiviral top hit drugs like Indinavir (ΔG= -9.8 kcal/Mol) and Nelfinavir (ΔG= -9 kcal/Mol) are binding near the center of S2 (Fig. 2). The similar pattern observed with antiparasitic drugs Ivermectin B1a/b (ΔG= -9.16/-8.86 kcal/Mol) (Fig. 4A-4B) and Vitamin D (ΔG= -5.52 kcal/Mol) (Fig. 4E-4F), whereas on the other hand, top hit antibiotic drugs did not show preferences at the S2 binding area of the Spike protein (Fig. 4). In this perspective, the above-mentioned drugs could possess the ability to physically inhibit the assembly of the Spike protein as well as the viral entry in a pan-CoV targeted manner.

Shuai and colleagues (Xia, Yan et al. 2019) showed that targeting the HR1 domain of Human-CoV Spike protein, which is in the area of S2, that showed increased binding with our docked drugs, significantly reduced the viral entry in mouse models. Targeting this specific domain with a small molecule could be a pan-CoV target as its conservation is high among the CoV species. Turning the attention to conserved areas of Spike protein and not to the hypervariable S1-CTD will lead to a drug that not only has great inhibitory potential but also will possess a broad target activity against multiple coronavirus infection.

Moreover, taking into consideration another spotlight region of the Spike protein, the Polybasic furin Cleavage Site (PCS), which located in the boundaries of S1 and S2, was under our observations (Andersen, Rambaut et al. 2020).Throughout our docked library only one molecule could bind into that area. Vancomycin which exhibited the lowest binding energy (ΔG = -10.2 kcal/Mol) from all our docked drug categories, is the only drug that binds within the region of PCS in an approximate position of amino acid range between 654 and 692. The functional activity is still unknown for SARS-CoV-2 PCS; pathogenicity and transmissibility need to be elucidated as well. Emerging research referred to PCS as an enhancer of cell-cell fusion, whereas cleavage of it results in intra-species infection; with a great example of MERS-CoV (Menachery, Dinnon et al. 2020). Thus, it is very important to evaluate drugs that can bind to that area of the Spike protein since it might result in wanting outcomes for the fight against SARS-CoV-2.

## Conclusion

In the present study, docking of common antiviral, antibiotics, antiparasitic, and other ligands such as flavonoids and vitamins to SARS-Cov-2 core proteins (Spike and RdRp) revealed the potential of their use for the treatment of COVID-19 during the current emergency. Drugs Indinavir, Nelfinavir, Vancomycin, Gliclazide, Ivermectin, Vitamin C, and Vitamin D, showed very high affinity, thus can be used as a potential therapeutic in the current pandemic. However, *in-vitro* and *in-vivo* clinical trials are needed to further evaluate the potential of these drugs alone or in combination for the treatment of COVID-19. In addition, the finding of most interacting motifs of Spike and RdRp, altogether with conservative motifs of SARS-CoV-2 Spike throughout the evolution gives a new layer of understanding of the computer-aided drug and antibody/vaccine designing for the treatment of COVID-19. Our study suggests there is a high possibility of mutation in the S1 region of the Spike protein, in contrary S2 remains more conservative during the evolution. Therefore, the use of a combination of conserved peptide sequences of Spike protein could be a strategy to develop an effective vaccine. Additionally, designing the new drugs selectively targeting these conservative motifs might help to find out a potential cure for the COVID-19. However, an *in-vitro/in-vivo* investigation will be needed for the further clinical use of the results of present study. Finally, designing of inhibitory peptides blocking the receptor-binding domain of Spike protein interacting with hemoglobin/heme that we reported in the current study, could help to control the hypoxia condition during the infection of SARS-Cov-2, which might significantly improve the survival of patients suffering from COVID-19.

**Acknowledgment**

LA would like to acknowledge Prof. Shiga Takashi (University of Tsukuba) and Dr. Carl (Chancellor, *UNICAL,* Curacao) for their support during the study. The authors do not have any grant support for the current research.

**Conflict of Interest**

The authors declare that there is no conflict of interest that could be perceived as prejudicial to the impartiality of the reported research.

**Author Contributions**

LA and TP developed the idea, designed the study, performed the experiments, data analysis, and contributed equally in the manuscript. SE, CM, and RKB contributed to the development of the idea, helped in the simulation, writing, and revision of the manuscript. All the authors read and approved the final manuscript for submission. The authors declare that there is no conflict of interest that could be perceived as prejudicial to the impartiality of the reported research.

**Abbreviations**

3Clpro : 3C-Like protease; ACE2 : Angiotensin-Converting Enzyme 2; COVID-19 : Coronavirus Disease 2019; DSV: Discovery Studio Visualizer; HE : Hemagglutinin-esterase dimer; Hgb : Hemoglobin; HR1 : Heptad Repeat 1; HR2 : Heptad Repeat 2; I-TASSER : Iterative Threading ASSEmbly Refinement; MEGA : Molecular Evolutionary Genetics Analysis; MERS : Middle East Respiratory Syndrome; PCS : Polybasic furin Cleavage Site; RdRp : RNA-directed RNA polymerase; S1-CTD : SpikeSubunit1-C Terminus Domain; S1-NTD : SpikeSubunit1-N Terminus Domain; S2 : SpikeSubunit2; SARS-CoV-2 : Severe Acute Respiratory Syndrome-CoronaVirus-2; URIs: Upper Respiratory Tract Infections; WHO : World Health Organization

**Figures**

A

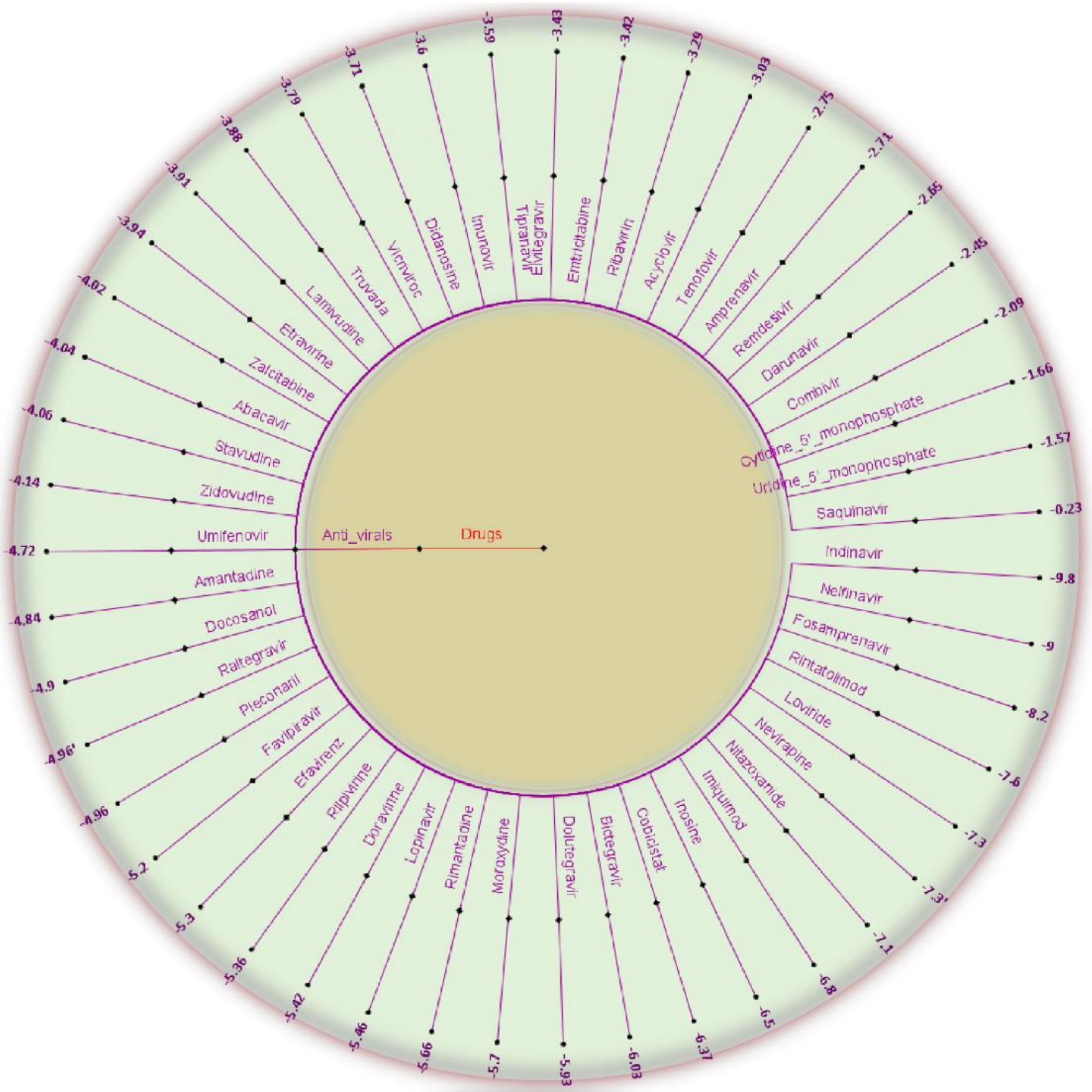

B

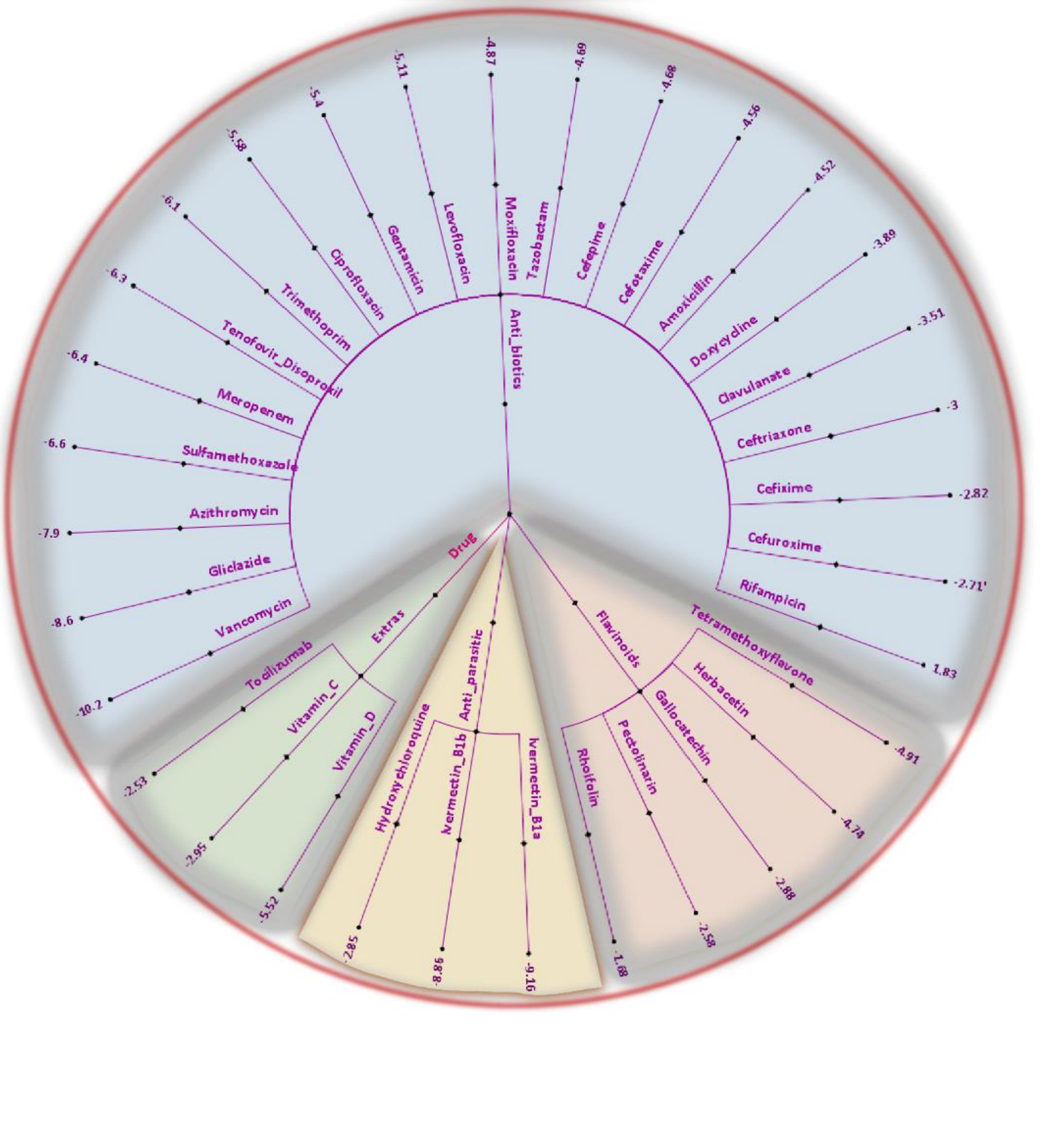

**Fig.1**: Circular tree represents the order of antiviral drugs (A) and antibiotics, antiparasitic, flavonoids, and extra ligands (B) based on the binding energy (ΔG in kcal/Mol).

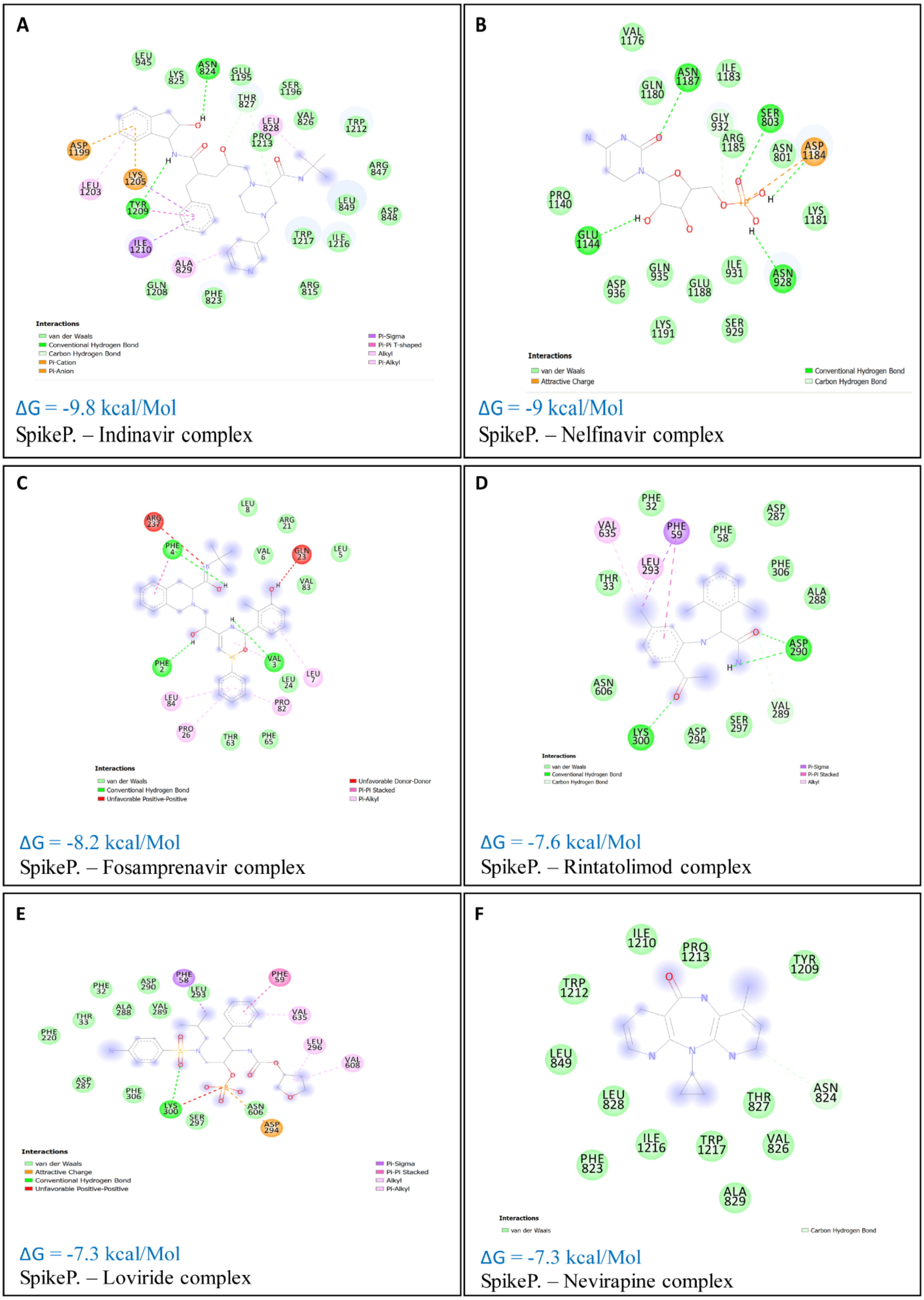

**Fig.2**: 2D interaction plot shows interface amino acid of Spike to top hit antiviral drugs Indinavir (A), Nelfinavir (B), Fosamprenavir (C), Ritatolimod (D), Loviride (E) and Nevirapine (F). ΔG is the binding energy of the drug ligand in kcal/Mol.

A

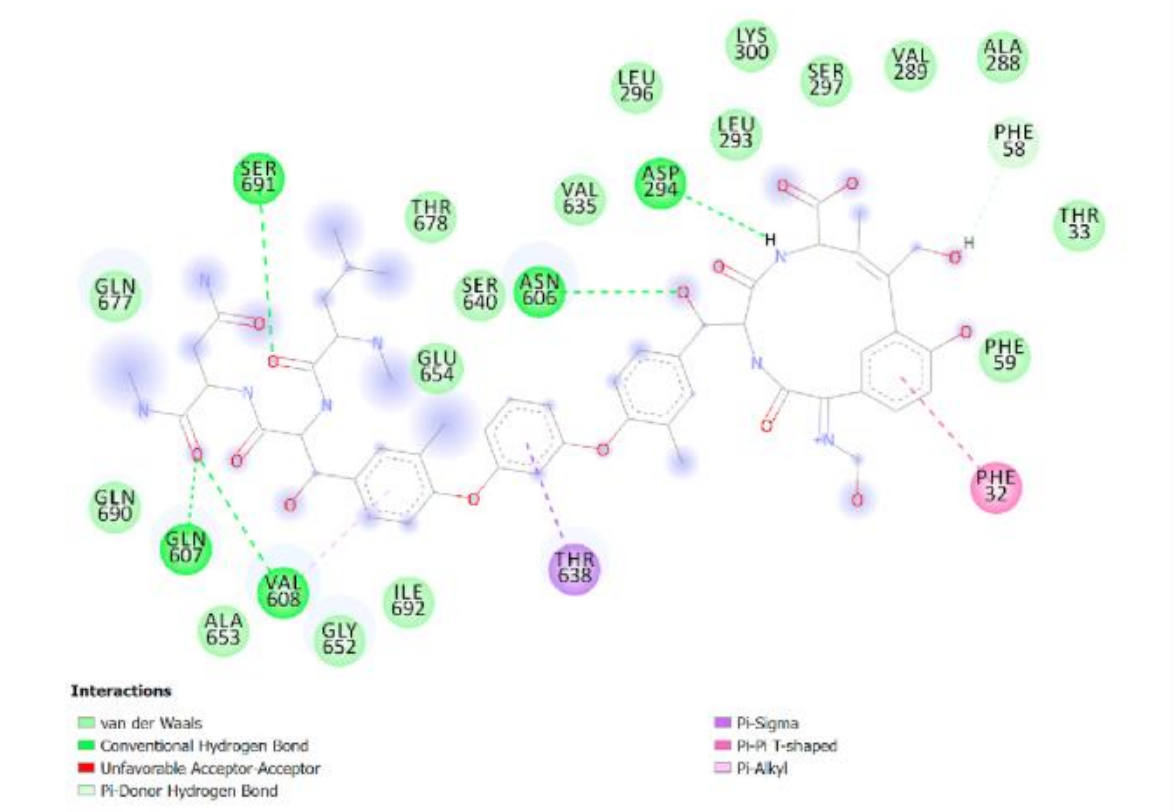

ΔG = -10.2 kcal/Mol
SpikeP. – Vancomycin complex

B

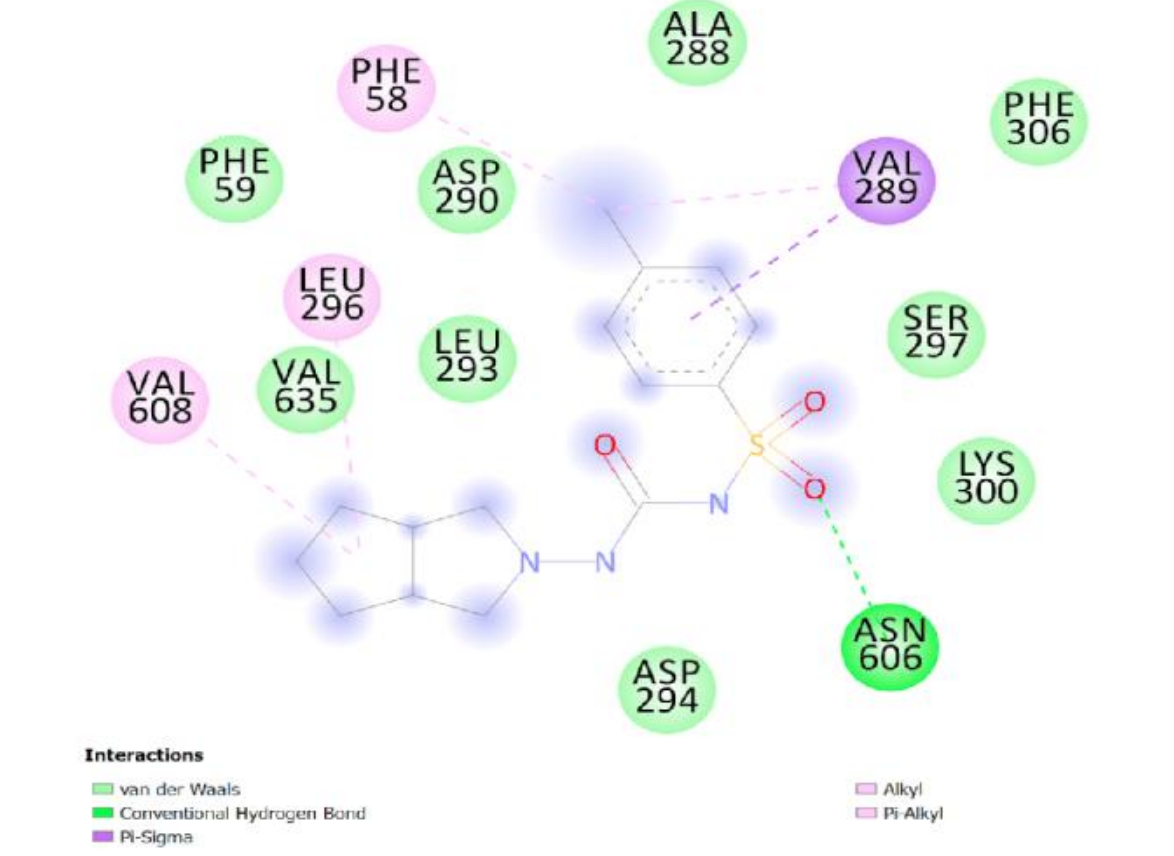

ΔG = -8.6 kcal/Mol
SpikeP. – Gliclazide complex

C

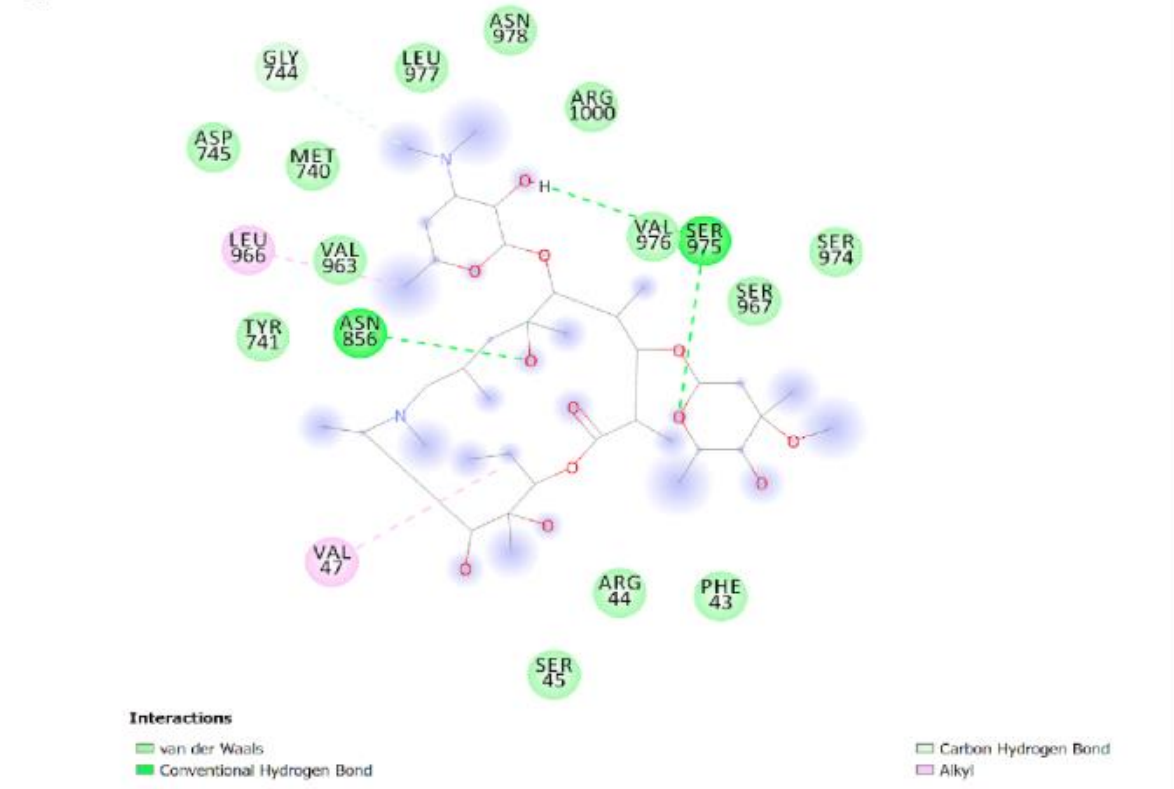

ΔG = -7.9 kcal/Mol
SpikeP. – Azythromycin complex

D

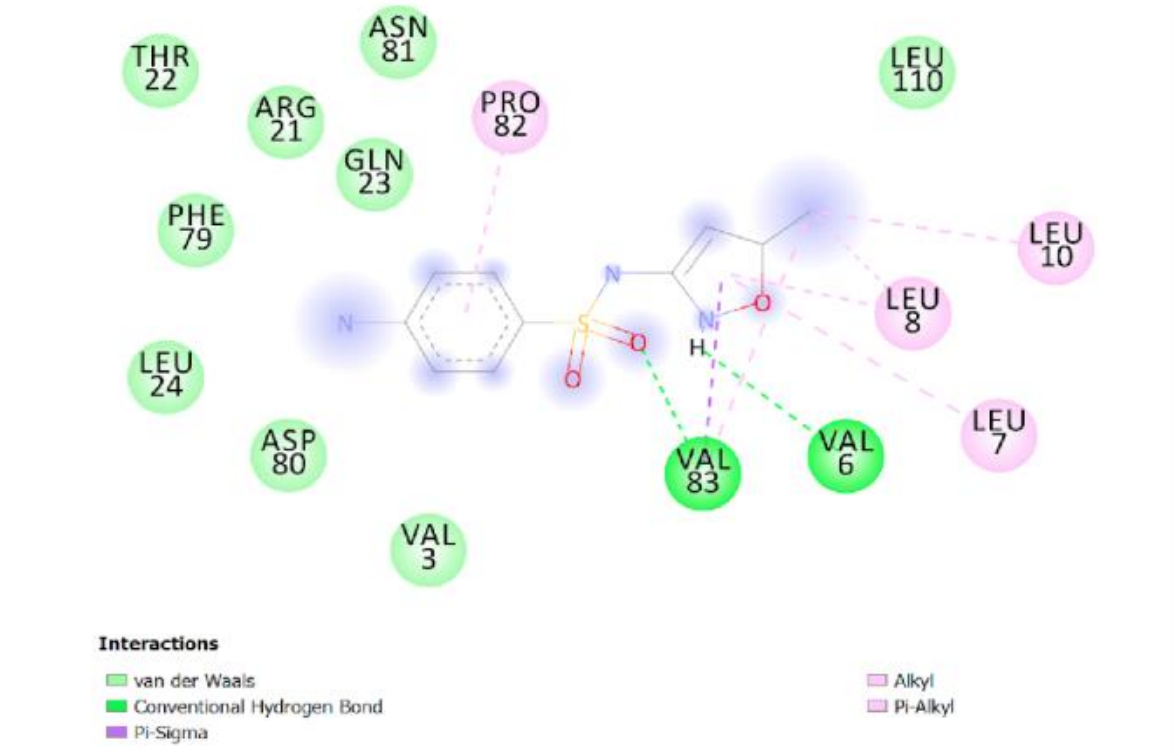

ΔG = -6.6 kcal/Mol
SpikeP. – Sulfamethoxazole complex

E

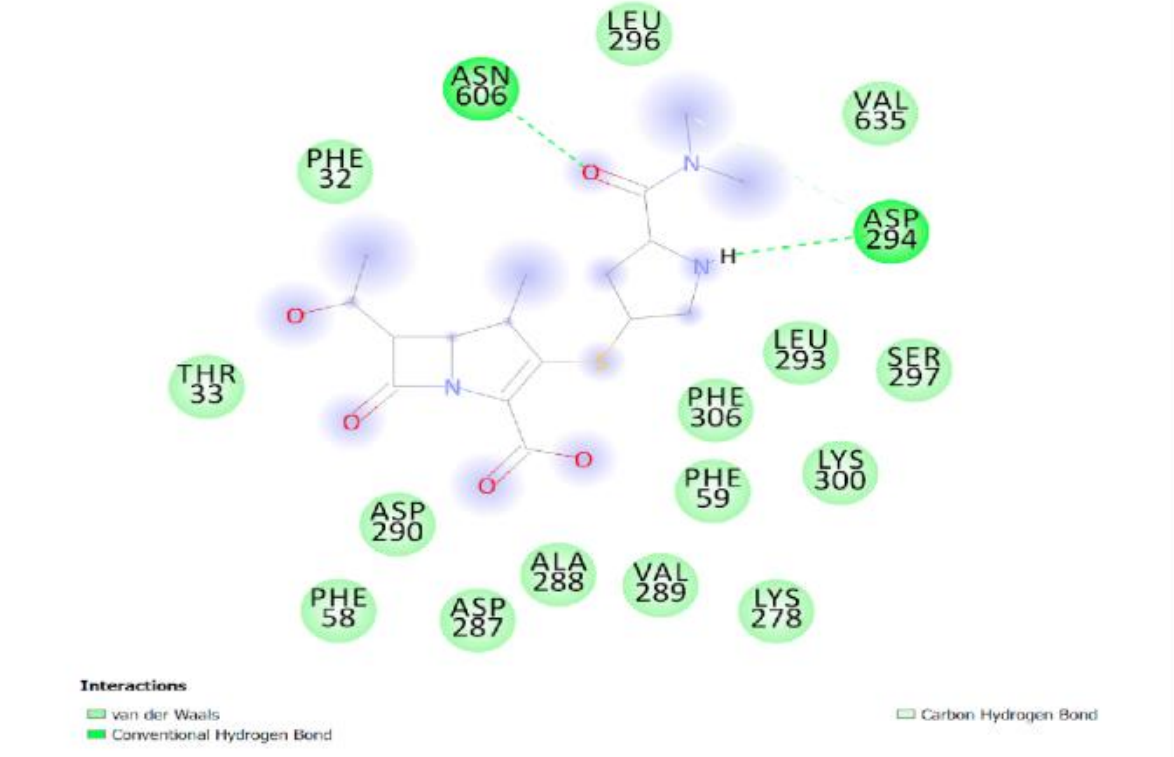

ΔG = -6.4 kcal/Mol
SpikeP. – Meropenem complex

F

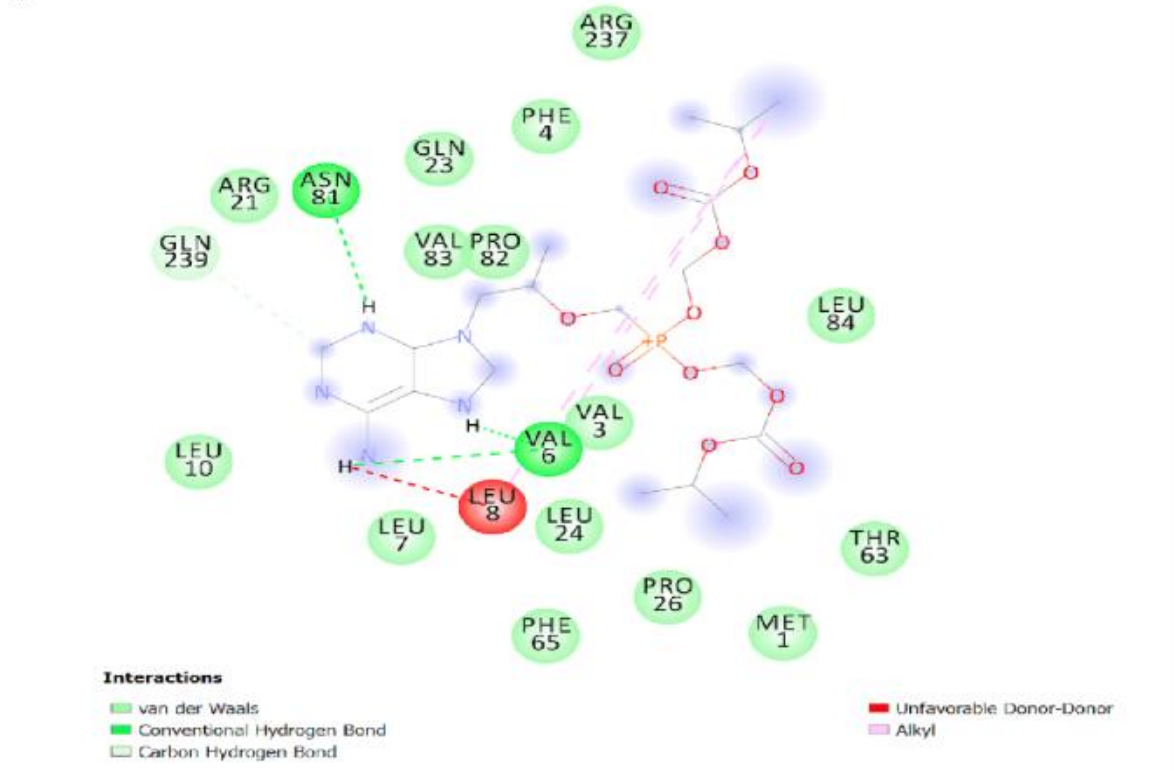

ΔG = -6.3 kcal/Mol
SpikeP. – Tenofovir disoproxil complex

**Fig.3**: 2D interaction plot shows interface amino acid of Spike to top hit antibiotics. Vancomycin (A), Gliclazide (B), Azithromycin (C), Sulfamethoxazole (D), Meropenem (E), and Tenofovir (F). $\Delta G$ is the binding energy of the drug ligand in kcal/Mol.

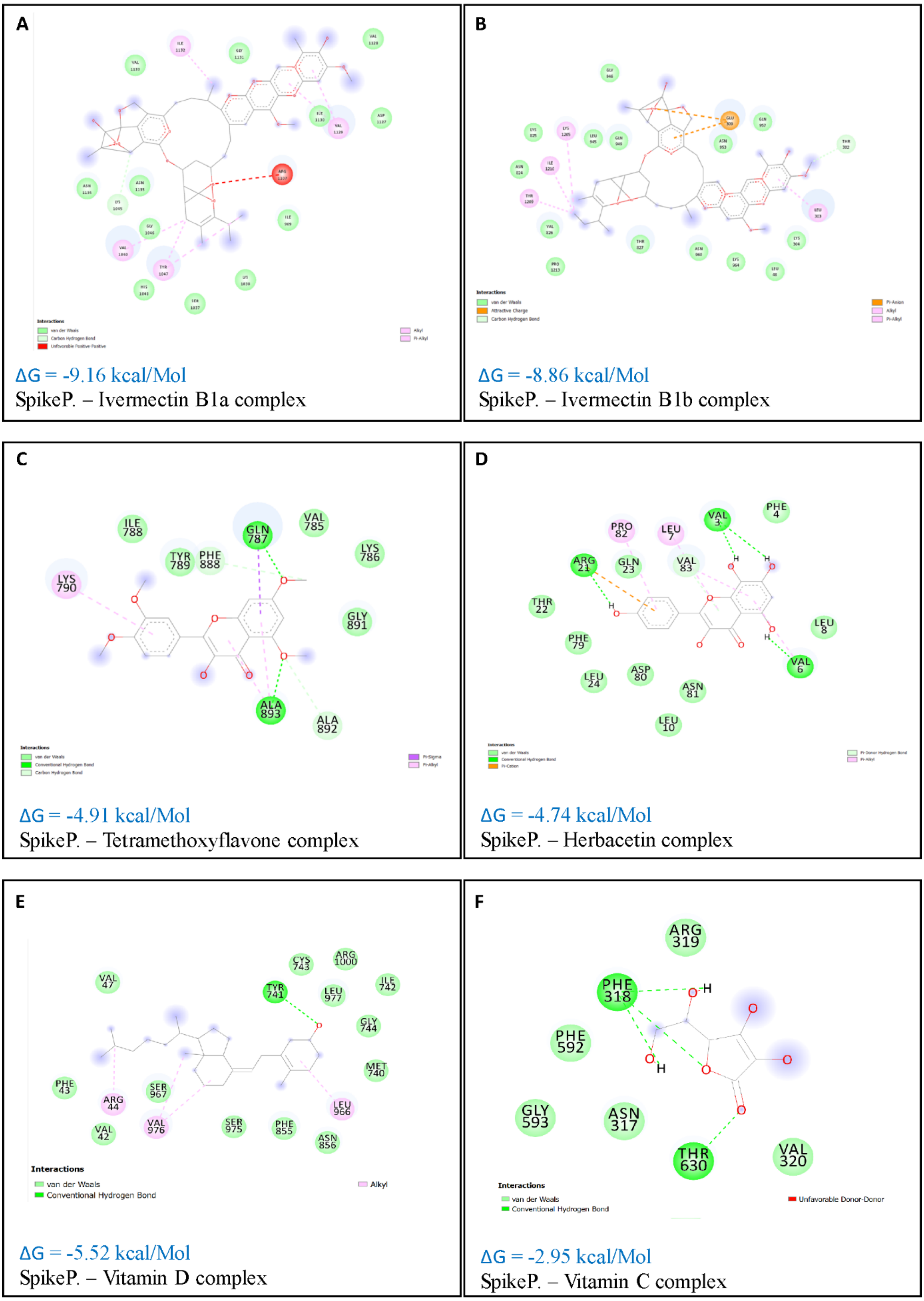

**Fig.4**: 2D interaction plot shows interface amino acid of Spike to top hit antiparasitic drugs (A and B), flavonoids (C and D), and vitamins (E and F). ΔG is the binding energy of the drug ligand in kcal/Mol.

A

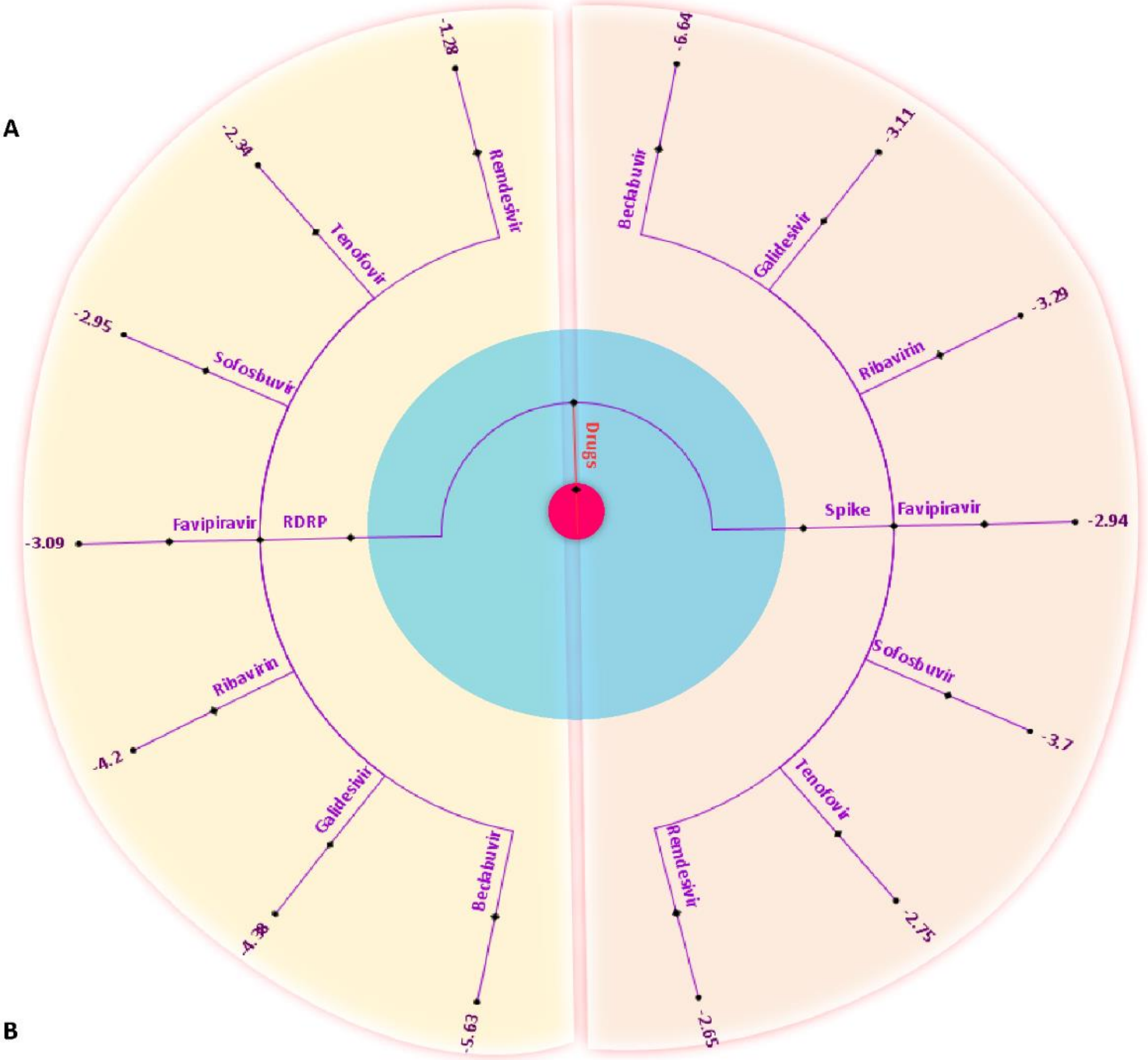

B

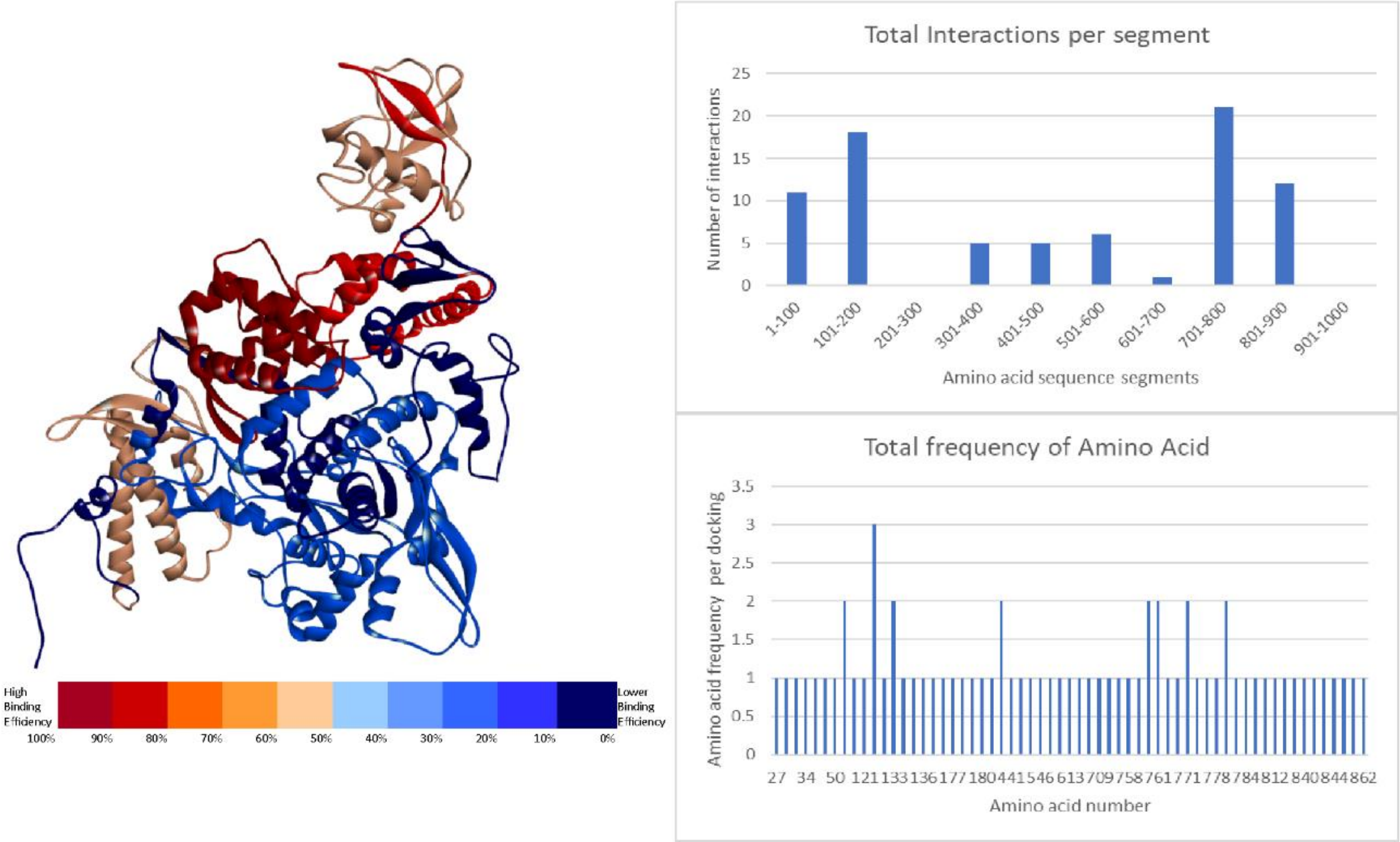

**Fig.5**: A graphical representation of the comparative analysis found in Table S4. Binding energies are organized here (A) are arranged from weakest to strongest based on the molecule it is interacting with. These lists of drugs and their binding energies are separated along the central axis based on the viral protein they are interacting with, namely RdRp on the left and the Spike protein on the right-hand side. Additionally, quantitative analysis and graphical representation of the locations at which the molecules are interacting with the SARS-CoV-2 RdRp (B) can be observed.

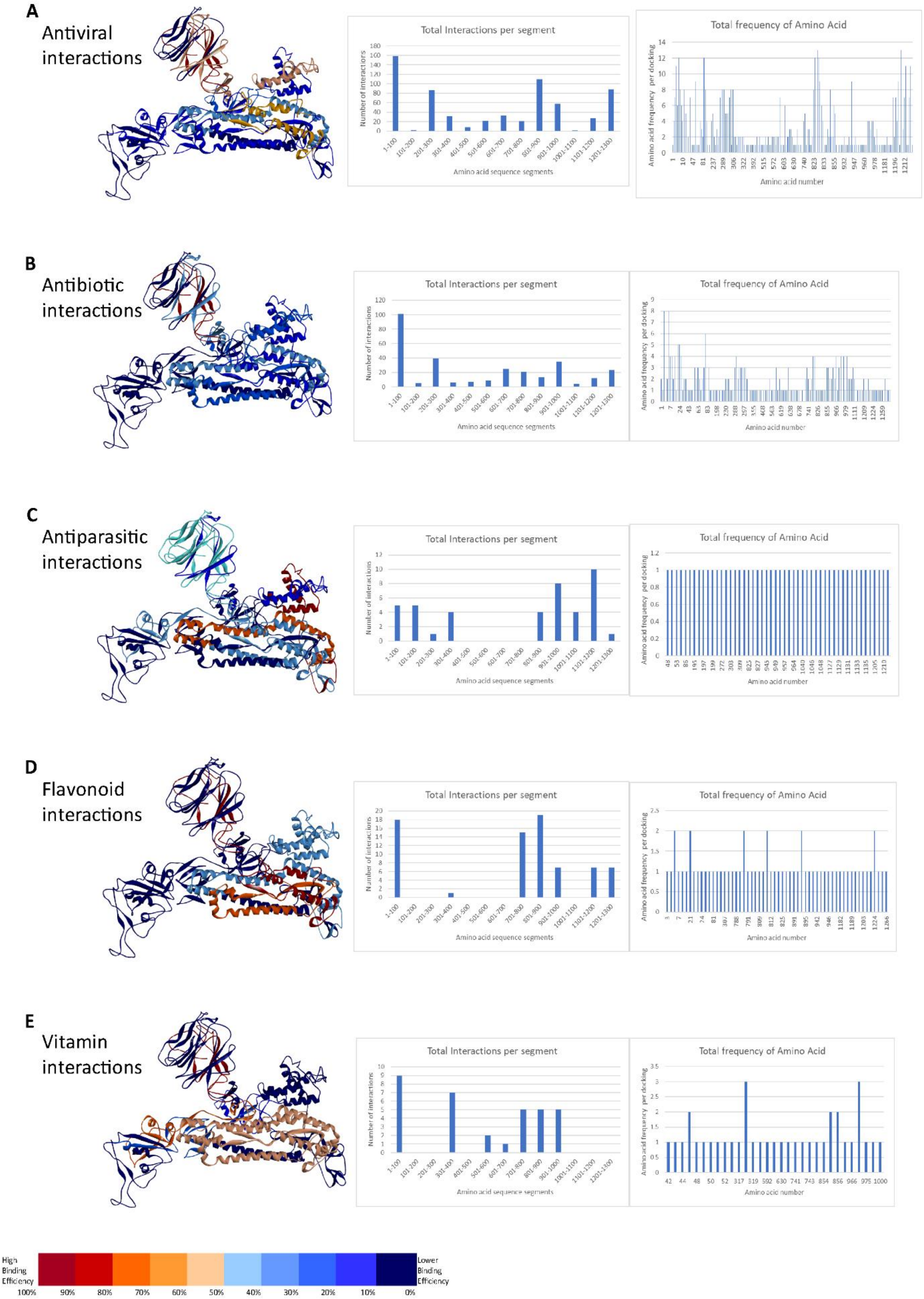

**Fig.6**: Heat map of the Spike protein represents the frequency of antiviral drug bindings against the 12 segments of the protein. Simulations suggest a preference of antiviral drugs to bind around the S1-NTD region and S2(A); were similar results observed in flavonoid interactions

(D). Docking of antibiotic drugs exhibits a preference in the S1-NTD region only (B). Antiparasitic drugs interestingly showed a higher binding frequency in S2 of the protein (C). From the perspective of the vitamins, they showed to bind in both S1-NTD/–CTD region and S2 (E).

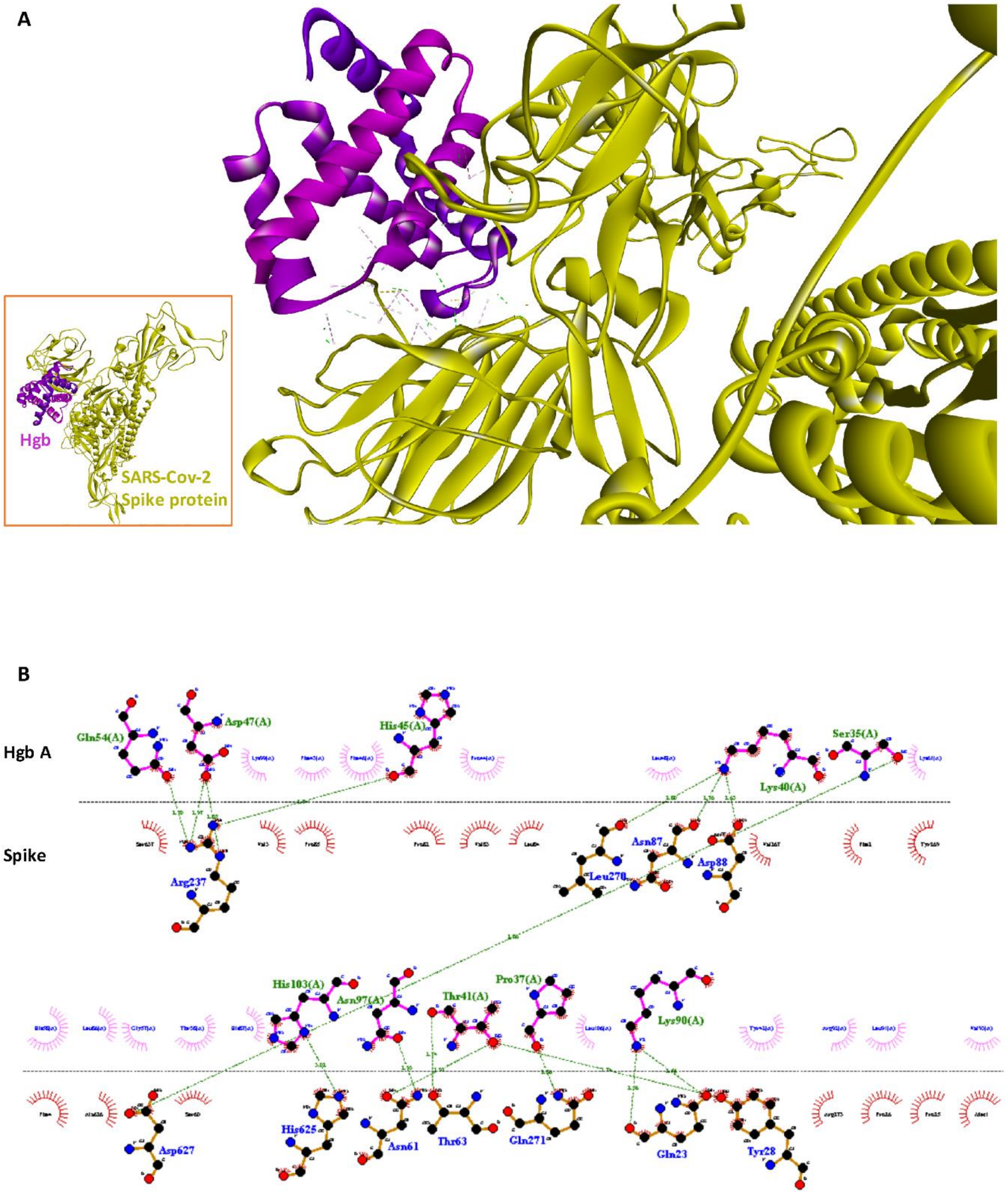

**Fig.7**: 3D representation of the interaction between the Spike protein (in yellow color) and hemoglobin monomer (in magenta color) (A). 2D interaction plot was formed via LigPlot. The interaction may occur at S1-NTD with the referred amino acids from each side. Hydrogen bonds are represented by the green dashed lines (B).

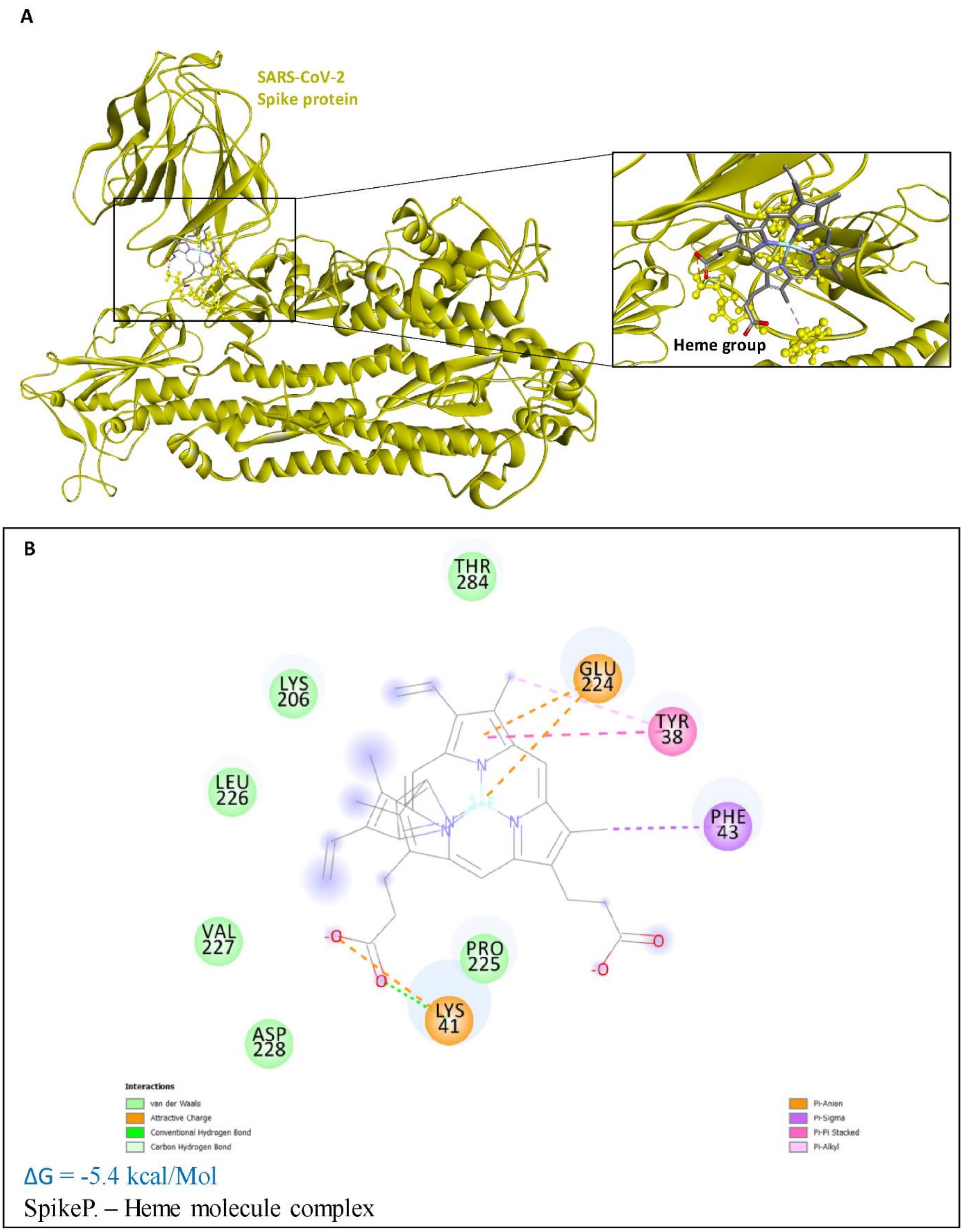

**Fig.8**: Spike protein (in yellow color) docked with hemoglobin heme group near the S1-NTD (A). 2D interaction plot of the interaction, whereas GLU224 seems to play a central role in this interaction via direct binding with the iron particle of heme. The binding energy of this reaction was calculated at -5.4 kcal/Mol (B).

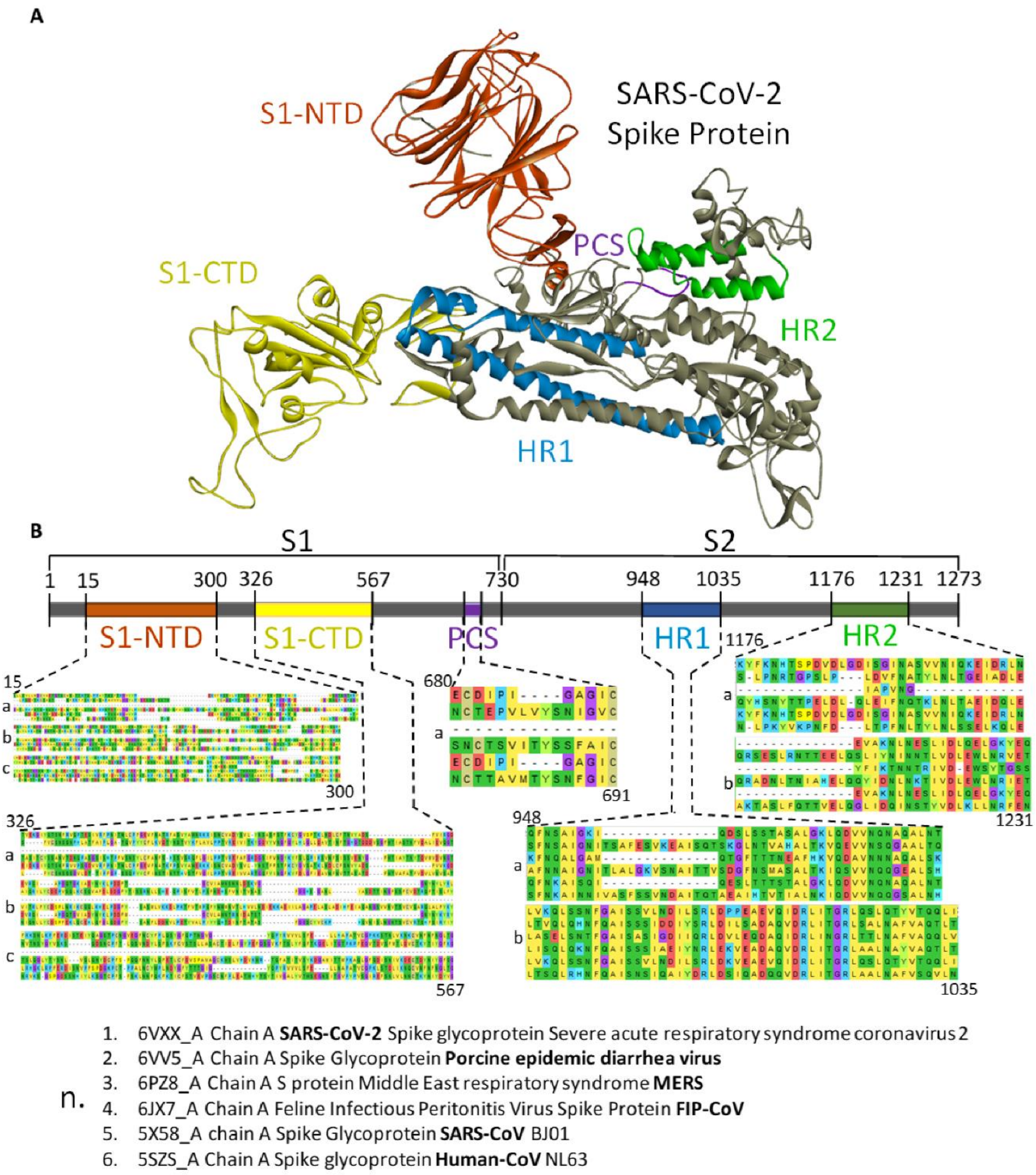

**Fig.9**: SARS-CoV-2 Spike protein 3D structure with the important domains depicted with colors; S1-N Terminal Domain (orange), S1-C Terminal Domain/ ACE2 Binding Domain (yellow), Polybasic Cleavage Site (purple), S2 Heptad repeat 1 (blue) and S2 heptad repeat 2 (green) (A). SARS-CoV-2 Spike protein 2D representation. Important domains are magnified, and their alignment sequences are shown beneath them. Each alignment box consists of 1 to 3 lanes (a., b., c.), whereas the starting amino acid is shown at the top left, and ending amino acid is shown at the bottom right of each alignment box. Each lane consists of 6 sequences subjected to the alignment (n.). Conservation exists in higher levels at S2-HR1 and HR2, moderate conservation is observed at S1-NTD and PCS, and at S1-CTD there are many hypervariable regions (B).

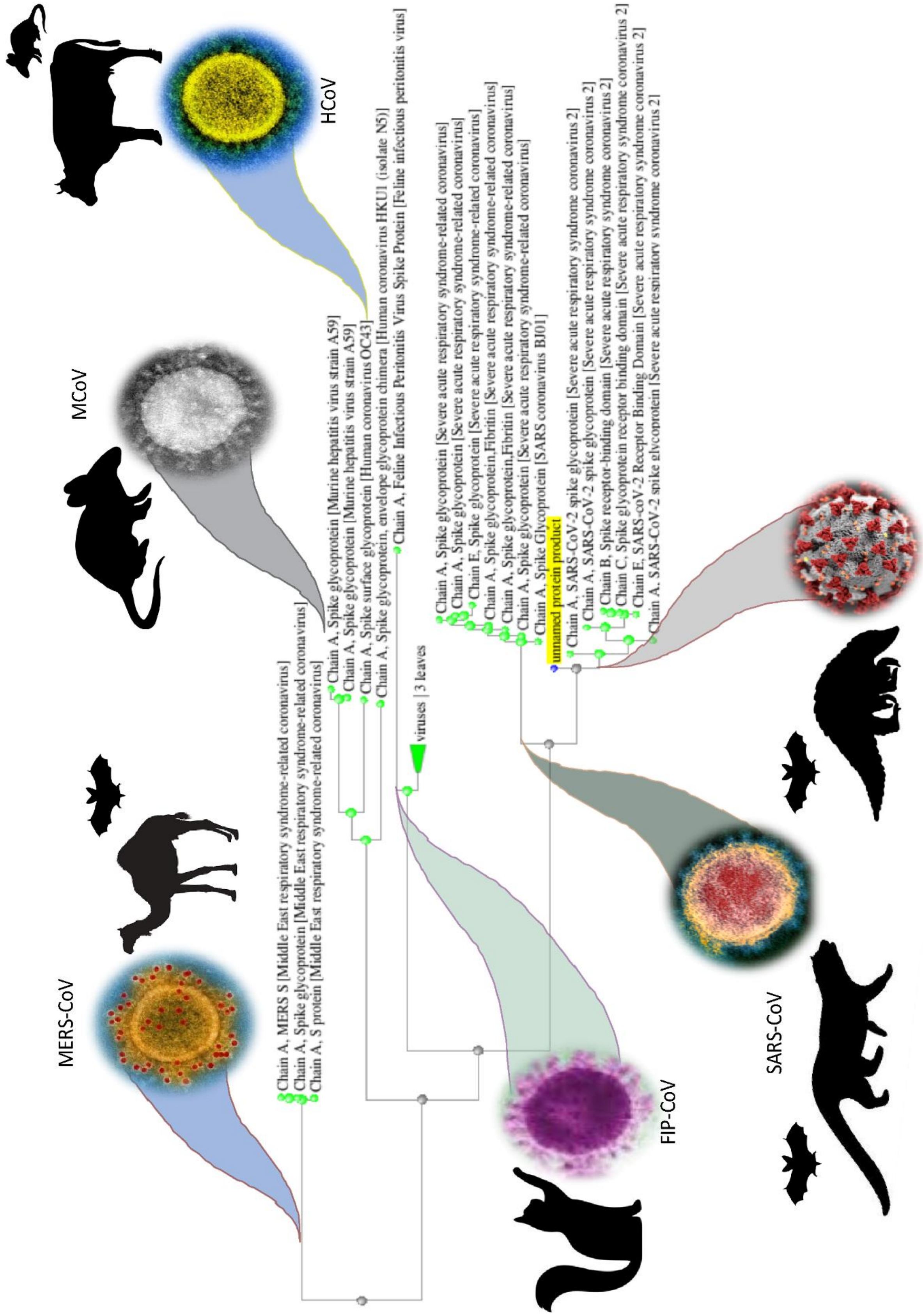

**Fig.10**: Phylogenetic tree of various coronavirus particles whose sequences align with SARS-CoV-2 Spike protein in the BLASTP server. Cartoon of the animal for each CoV species represents the carrier of the disease, whereas the secondary smaller animal represents the species that have been originated from if it is known.

# Supplementary Information for

**Viroinformatics-based investigation of SARS-CoV-2 core proteins for potential therapeutic targets**

***Lokesh Agrawal[1,2,3*], Thanasis Poullikkas[4,5], Scott Eisenhower[4,6], Carlo Monsanto[7], Ranjith Kumar Bakku[8,9]***

[1]*Universidad Integral del Caribe y América Latina, Kaminda Cas Grandi #79, Curaçao*
[2]*Graduate School of comprehensive human sciences, University of Tsukuba, 1-1-1 Tennodai, Tsukuba 305-8577, Japan*
[3]*Molecular Neuroscience Unit, Okinawa Institute of Science and Technology Graduate University, Kunigami-gun, Okinawa 904-0412, Japan*
[4]*Human Biology, School of Integrative and Global Majors, University of Tsukuba, 1-1-1 Tennodai, Tsukuba 305-0006, Japan*
[5]*Department of Experimental Pathology, Faculty of Medicine, University of Tsukuba, 2-1-1 Tennodai, Tsukuba 305-8576, Japan*
[6]*Department of Infection Biology, Faculty of Medicine, University of Tsukuba, 1-1-1 Tennodai, Tsukuba 305-8575, Japan*
[7]*Research Workgroup, Ronin Institute, 127 Haddon Place, Montclair, NJ 07043-2314 USA*
[8]*Faculty of Engineering Information and Systems, Dept. of Computer Science, University of Tsukuba, 1-1-1 Tennodai, Tsukuba, Ibaraki, 305-8572, Japan*
[9]*Tsukuba Life Science Innovation Program (TLSI), University of Tsukuba, 1-1-1 Tennodai, Tsukuba, Ibaraki, 305-8572, Japan*

***Corresponding Author**
*Email: lokesh.agrawal@unical.university*
*Mobile: +81-8094427502*

**This PDF file includes:**

Supplementary text
Figures S1 to S5
Tables S1 to S4

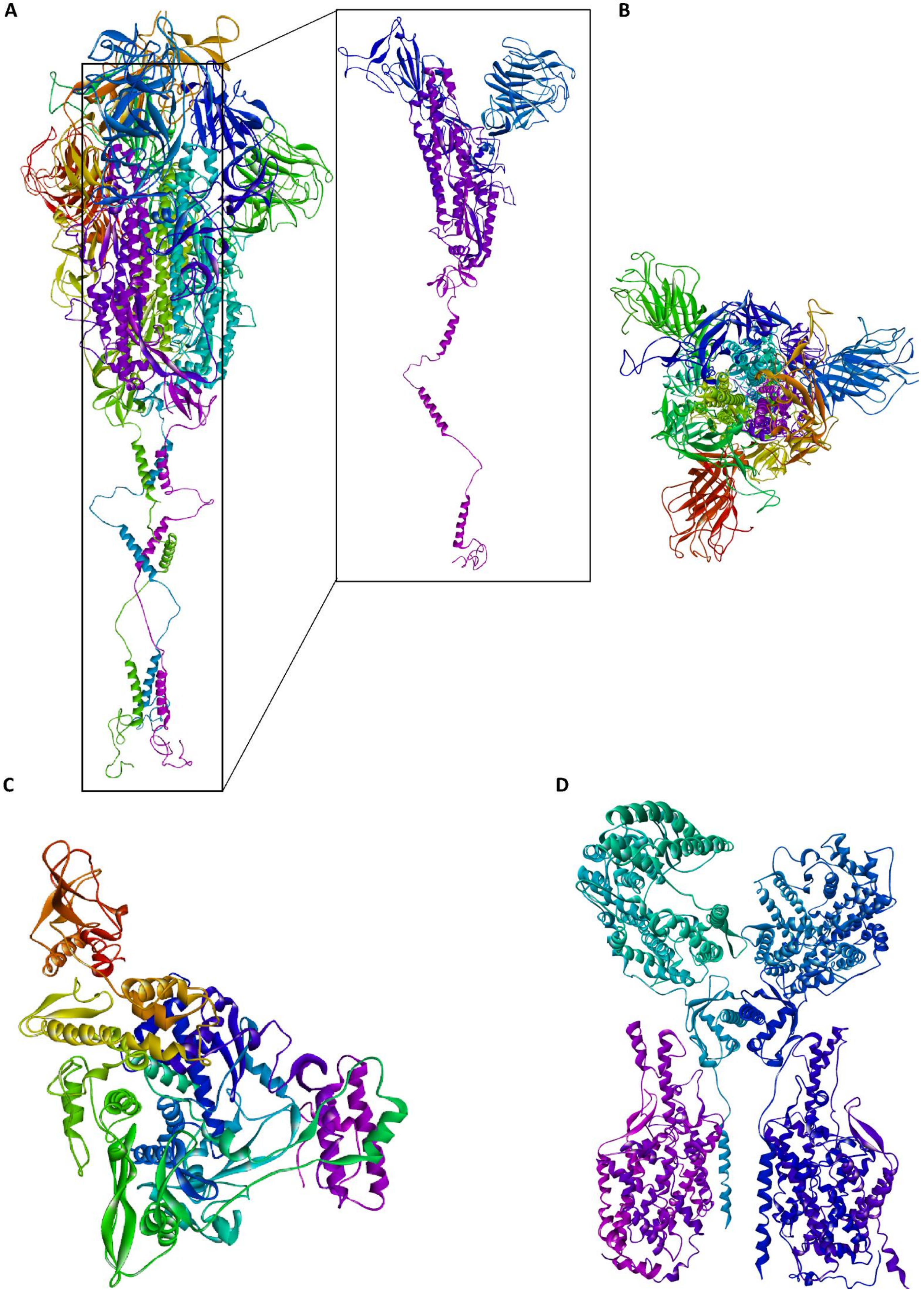
A
B
C
D

**Fig. S1**: 3D protein structure of the SARS-CoV-19 Spike protein in its trimeric form. Each monomer is colored with a different color (magenta, blue, and green) (A). Top view of the trimeric Spike protein (B). 3D protein structure of SARS-CoV-19 RdRp (C). 3D protein structure of a dimer of human ACE2 receptor (D).

**A**

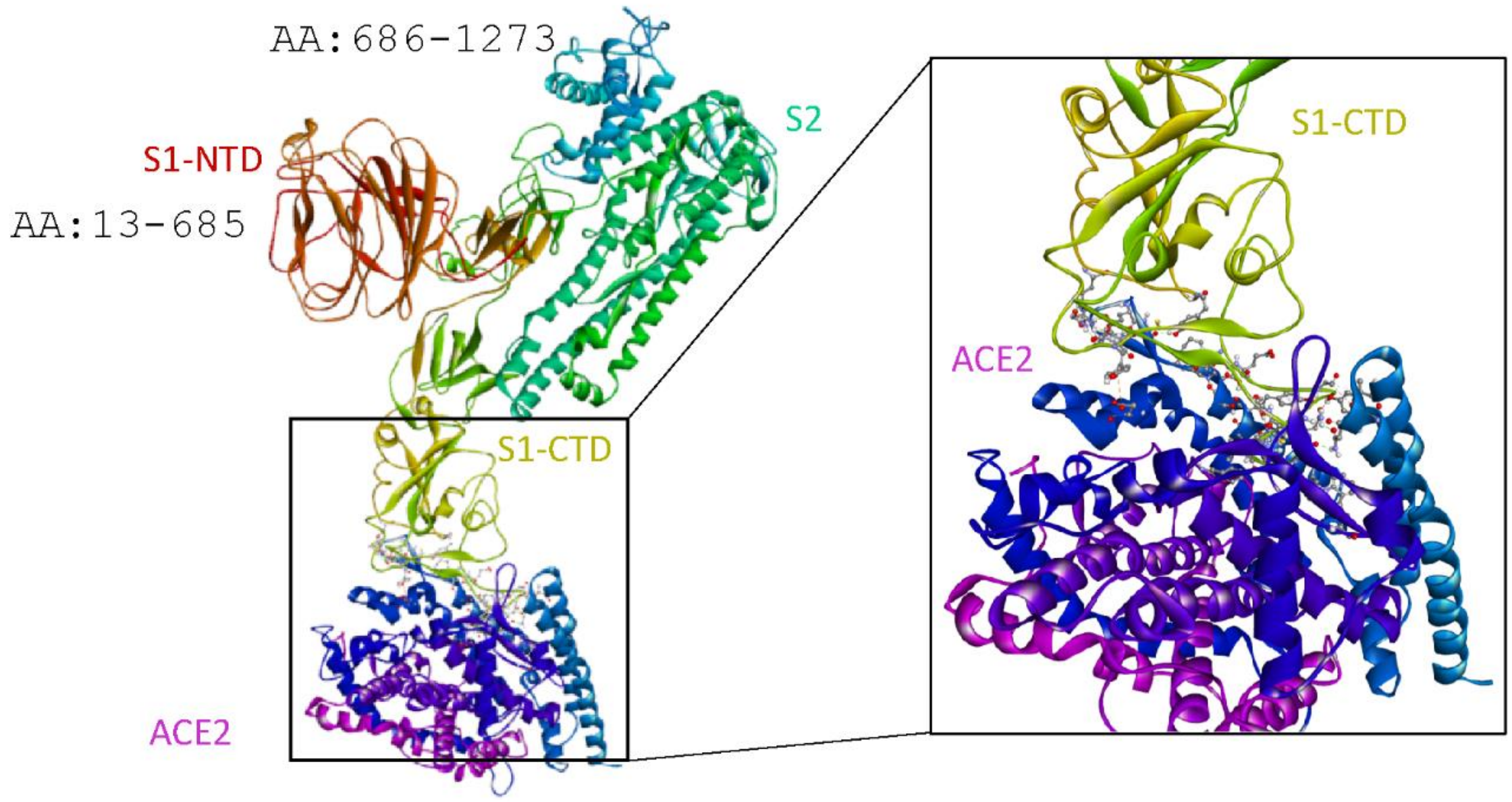

**B**

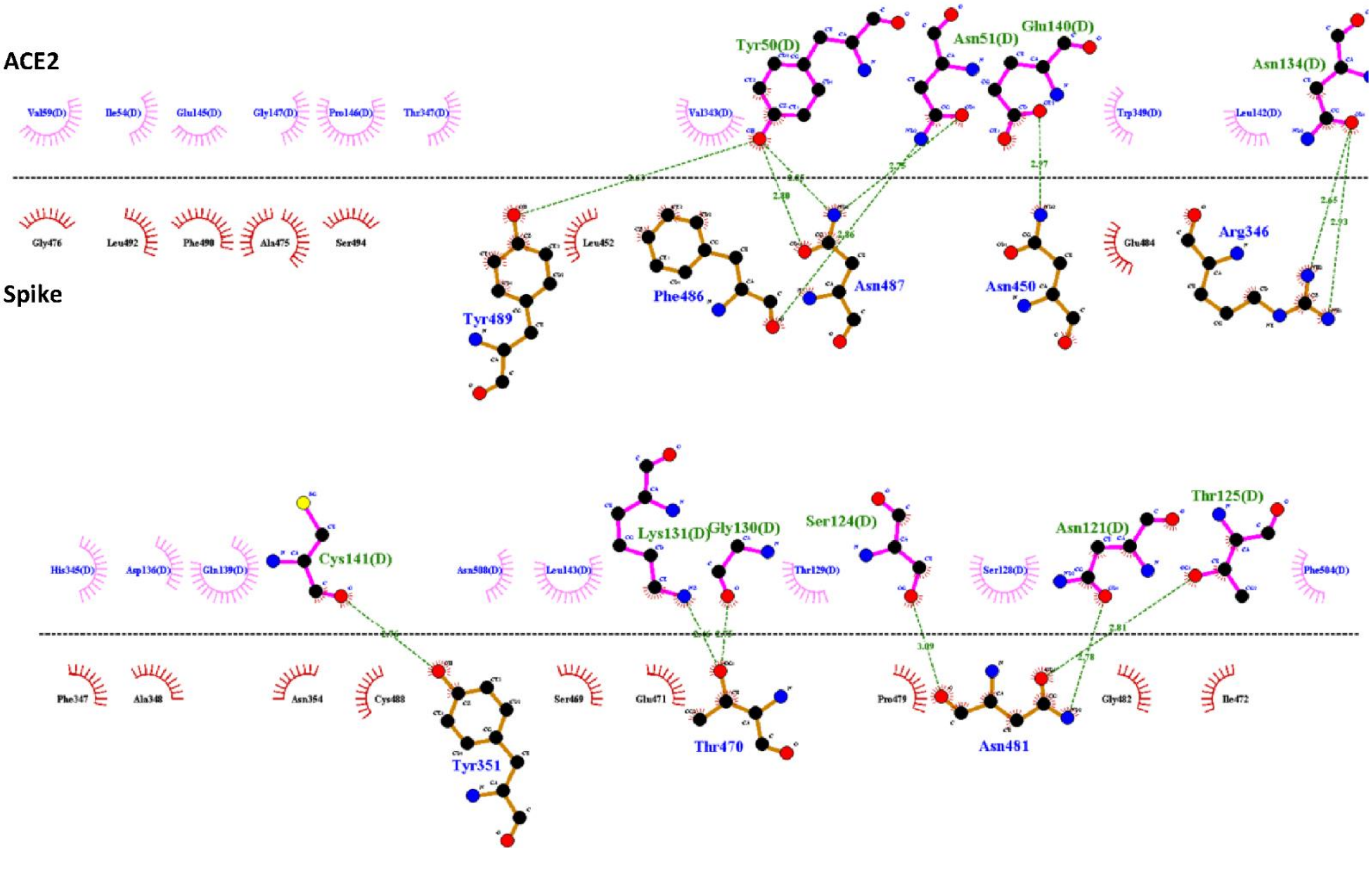

**Fig. S2**: Interaction of the trimeric SARS-CoV-19 Spike protein with the dimeric ACE2 receptor using the ClasPro server. Interaction is happening at the S1-CTD of the Spike protein (A). Interacting amino acids of each protein responsible for the interaction visualized by Ligplot. Green dashed lines between the opposing amino acids represent hydrogen bonds, whereas arches represent other forces (i.e. Van der Waals forces) (B).

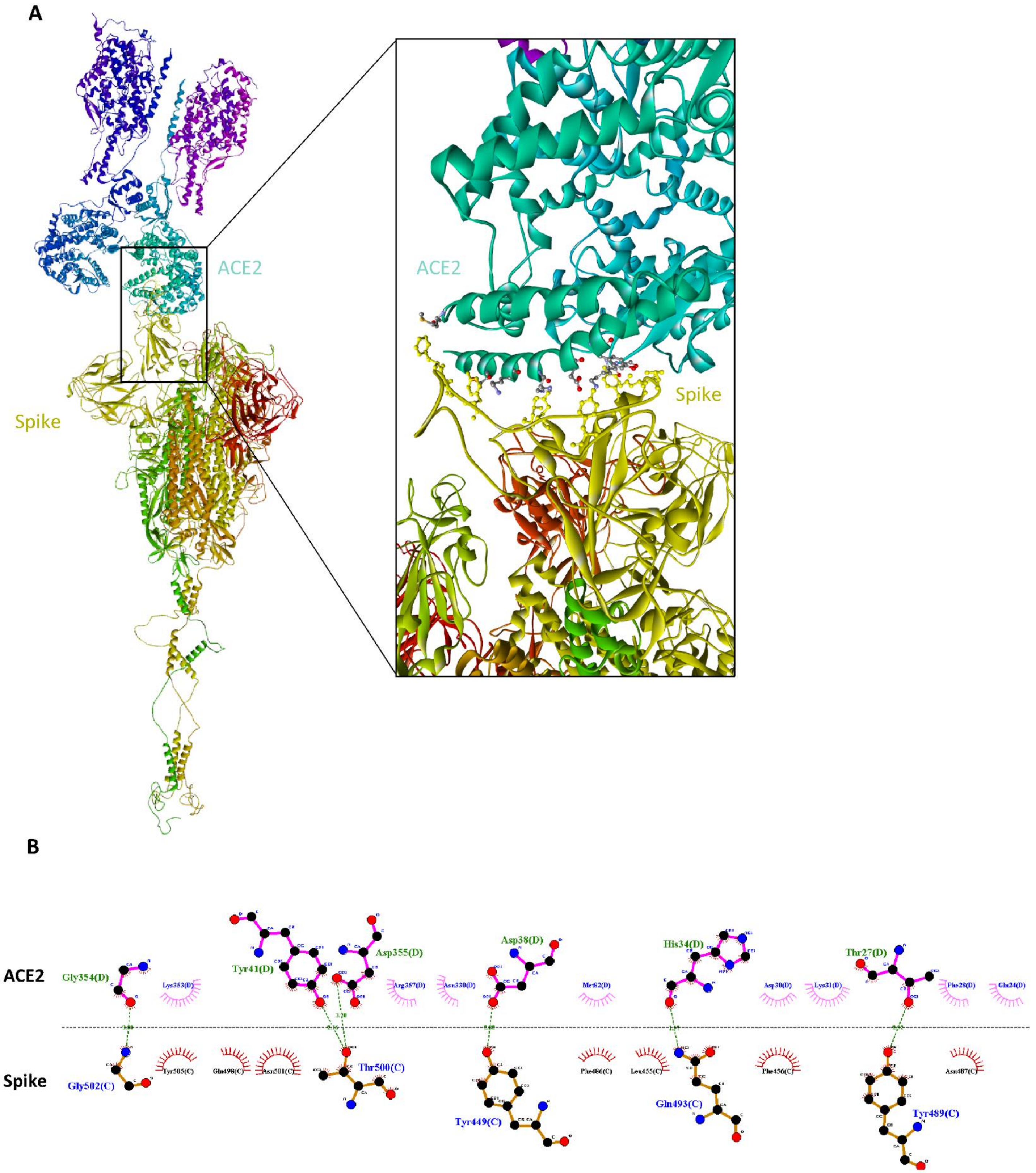

**Fig. S3**: Interaction of the monomeric SARS-CoV-19 Spike protein with the monomeric ACE2 receptor using the ClasPro server. Interaction is happening at the S1-CTD of the spike protein (A). Interacting amino acids of each protein responsible for the interaction visualized by Ligplot. Green dashed lines between the opposing amino acids represent hydrogen bonds, whereas arches represent other forces (i.e. Van der Waals forces) (B).

A

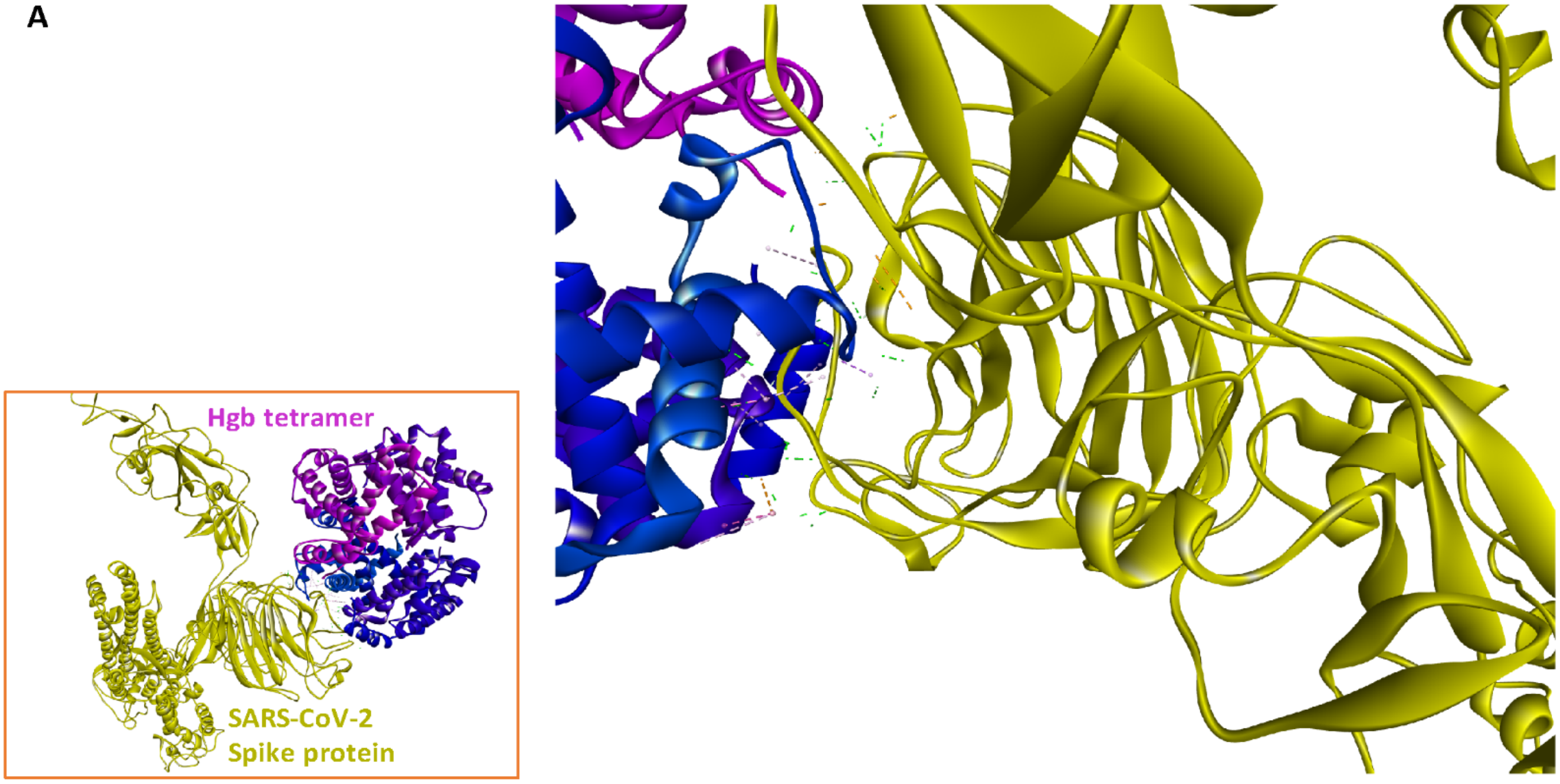

Hgb tetramer
SARS-CoV-2
Spike protein

B

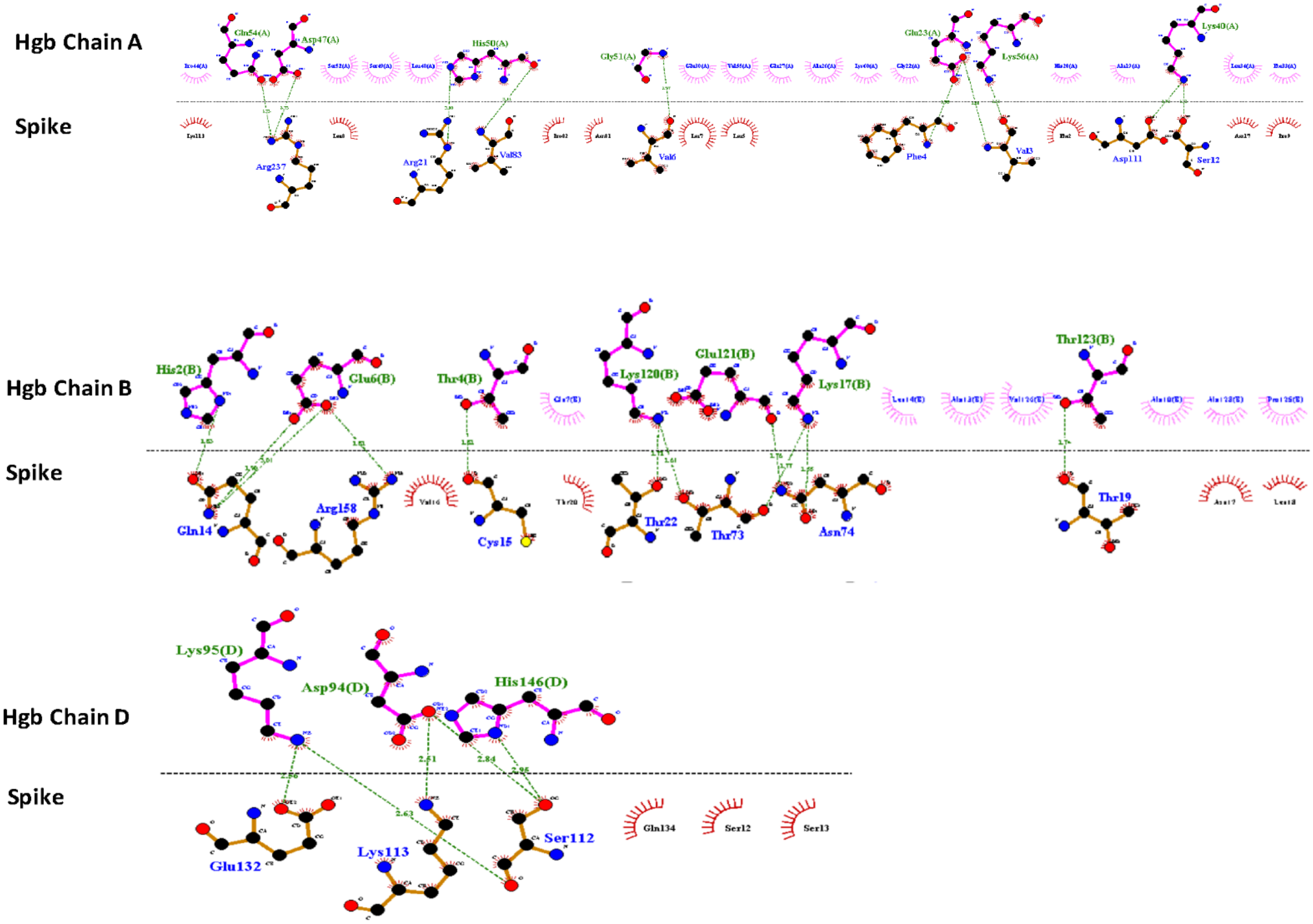

**Fig. S4**: 3D representation of the interaction between the Spike protein (in yellow color) and hemoglobin tetramer (in magenta color) (A). 2D interaction plot was formed via LigPlot. The interaction may occur at S1-NTD with the referred amino acids from each side. In green dashed lines represent hydrogen bonds between the interactive amino acids (B).

**A**

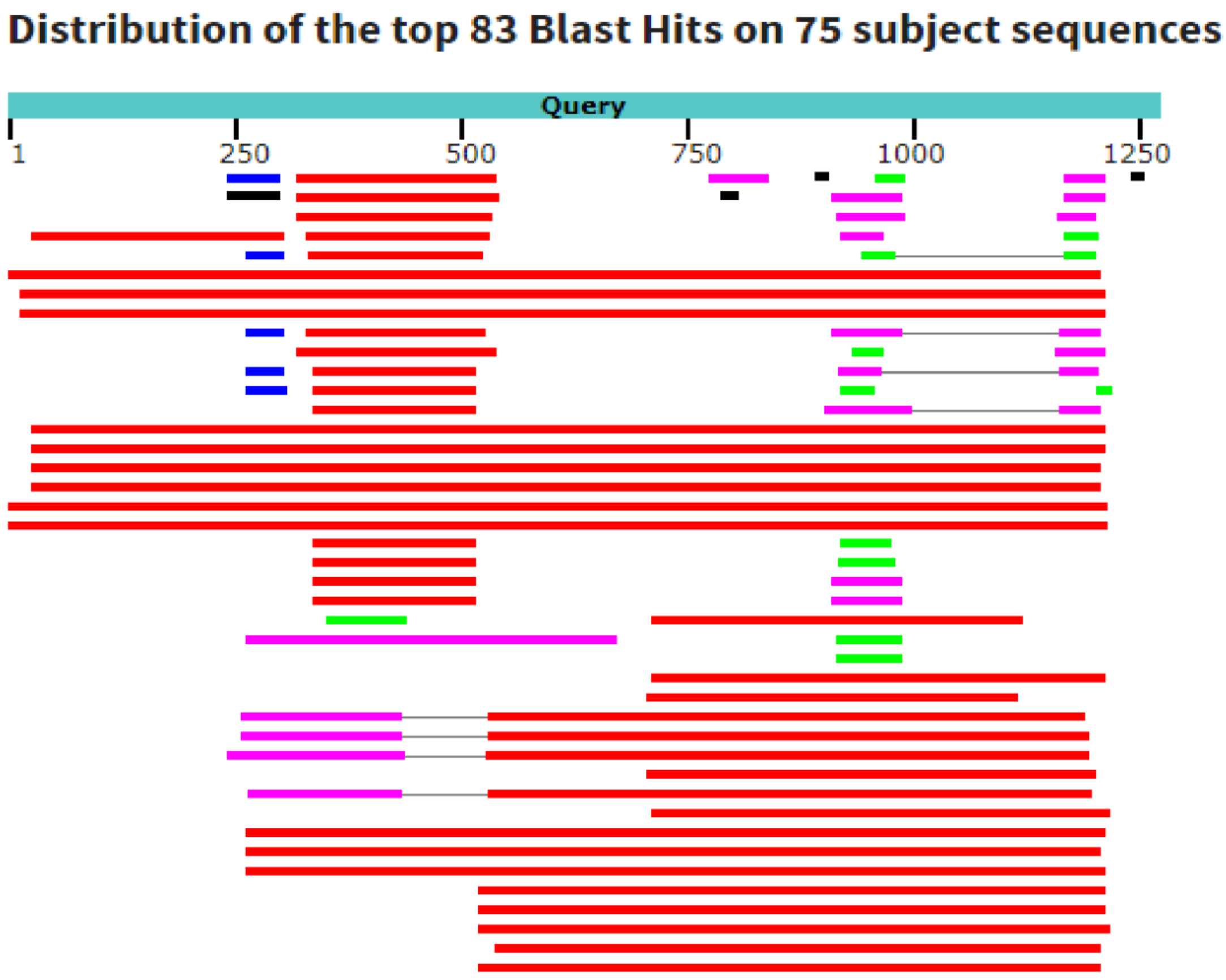

**B**

| Taxonomy | Number of hits | Number of Organisms | Description |
|---|---|---|---|
| ⊟Orthocoronavirinae | 300 | 22 | |
| . ⊟Betacoronavirus | 272 | 14 | |
| . . ⊟Sarbecovirus | 147 | 6 | |
| . . . ⊟Severe acute respiratory syndrome-related coronavirus | 106 | 6 | Severe acute respiratory syndrome-related coronavirus hits |
| . . . . Severe acute respiratory syndrome coronavirus 2 | 25 | 1 | Severe acute respiratory syndrome coronavirus 2 hits |
| . . . . SARS coronavirus BJ01 | 6 | 1 | SARS coronavirus BJ01 hits |
| . . . . SARS coronavirus Urbani | 2 | 1 | SARS coronavirus Urbani hits |
| . . . . SARS coronavirus PUMC03 | 6 | 1 | SARS coronavirus PUMC03 hits |
| . . . . SARS coronavirus FRA | 2 | 1 | SARS coronavirus FRA hits |
| . . ⊟Merbecovirus | 100 | 2 | |
| . . . ⊟Middle East respiratory syndrome-related coronavirus | 86 | 2 | Middle East respiratory syndrome-related coronavirus hits |
| . . . . Human betacoronavirus 2c EMC/2012 | 14 | 1 | Human betacoronavirus 2c EMC/2012 hits |
| . . ⊟Embecovirus | 24 | 5 | |
| . . . Murine hepatitis virus strain A59 | 13 | 1 | Murine hepatitis virus strain A59 hits |
| . . . ⊟Betacoronavirus 1 | 8 | 3 | |
| . . . . Human coronavirus OC43 | 6 | 1 | Human coronavirus OC43 hits |
| . . . . Bovine coronavirus | 1 | 1 | Bovine coronavirus hits |
| . . . . Dromedary camel coronavirus HKU23 | 1 | 1 | Dromedary camel coronavirus HKU23 hits |
| . . . Human coronavirus HKU1 (isolate N5) | 3 | 1 | Human coronavirus HKU1 (isolate N5) hits |
| . . Rousettus bat coronavirus HKU9 | 1 | 1 | Rousettus bat coronavirus HKU9 hits |
| . ⊟Alphacoronavirus | 19 | 5 | |
| . . Human coronavirus 229E | 7 | 1 | Human coronavirus 229E hits |
| . . Feline infectious peritonitis virus | 3 | 1 | Feline infectious peritonitis virus hits |
| . . Human coronavirus NL63 | 3 | 1 | Human coronavirus NL63 hits |
| . . ⊟Pedacovirus | 6 | 2 | |
| . . . ⊟Porcine epidemic diarrhea virus | 3 | 2 | Porcine epidemic diarrhea virus hits |
| . . . . Porcine epidemic diarrhea virus CV777 | 3 | 1 | Porcine epidemic diarrhea virus CV777 hits |
| . Infectious bronchitis virus | 3 | 1 | Infectious bronchitis virus hits |
| . ⊟Porcine coronavirus HKU15 | 6 | 2 | |
| . . Deltacoronavirus PDCoV/USA/Ohio137/2014 | 3 | 1 | Deltacoronavirus PDCoV/USA/Ohio137/2014 hits |
| . . Porcine deltacoronavirus | 3 | 1 | Porcine deltacoronavirus hits |

**Fig. S5**: Schematic represents a graphic summary of the alignment result in the BLASTP server. Query represents the SARS-CoV-2 Spike protein sequence aligned with 83 BLAST hits with a color grading system as the alignment score scale (black <40, blue 40-50, green 50-80, magenta 80-200 and red ≥200) (A). The table represents a taxonomical lineage of organisms possessing similar proteins as SARS-CoV-2 Spike protein. The number of hits represents the number of similar sequences compared with our query sequence. Orthocoronavirinae exhibits the highest number of hits followed by Betacoronavirus and Sarbecovirus (B).

**Table S1:** List of the ten top hits of antiviral drugs against SARS-CoV-2 Spike protein by order of lowest to highest binding energy value. Indinavir seems to exhibit the lowest binding energy (-9.8 kcal/Mol) when docked with Spike protein, whereas Cobicistat exhibits the highest binding energy (-6.37 kcal/Mol).

| | **Antivirals** | **Binding energy (kcal/Mol)** | **2D interaction** | **Interface amino acids** | **Type of interactions** |
|---|---|---|---|---|---|
| 1 | Indinavir (C36H47N5O4), 5362440 | -9.8 | 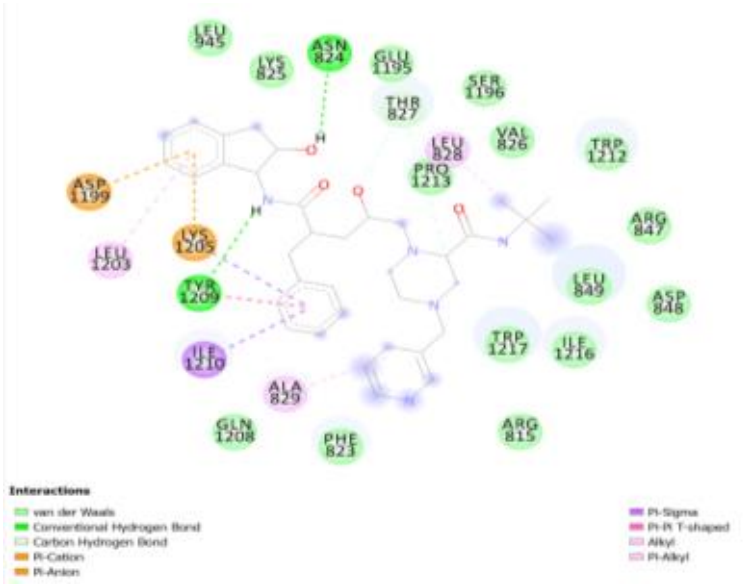 | Ala, 2Arg, Asn, 2Asp, Gln, Glu, 2Ile, 4Leu, 2Lys, Phe, Pro, Ser, Thr, 2Trp, Tyr | 1Alk, 1CHB, 2HB, 1Pi-An, 3Pi-Alk, 1Pi-Cat, 2Pi-Sig,1Pi-Pi-T, 14VW |
| 2 | Nelfinavir (C32H45N3O4S), 64143 | -9 | 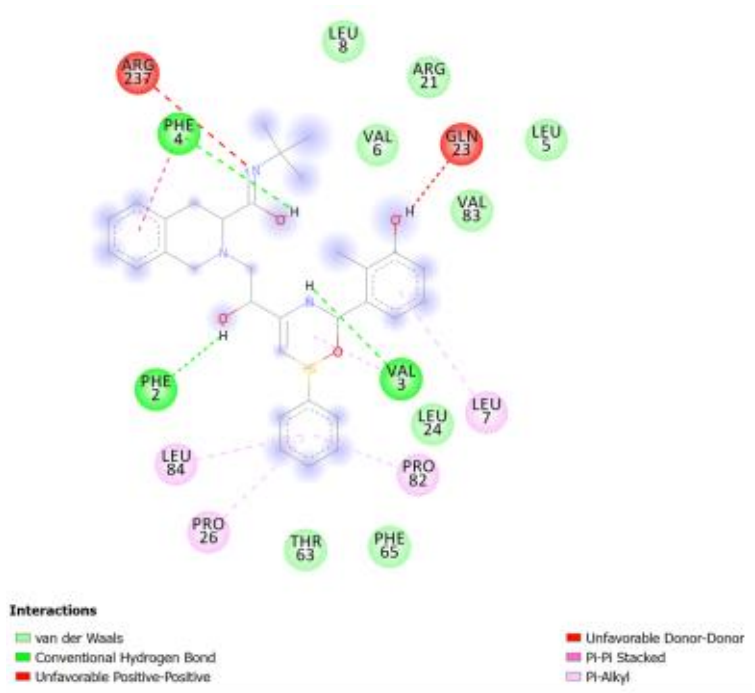 | 2Arg, Gln, 5Leu, 3Phe, 2Pro, Thr, 3Val | 3HB, 5Pi-Alk, Pi-Pi-Stk, 7VW, 1UDD, 1UPP |
| 3 | Fosamprenavir (C25H36N3O9PS), 131536 | -8.2 | 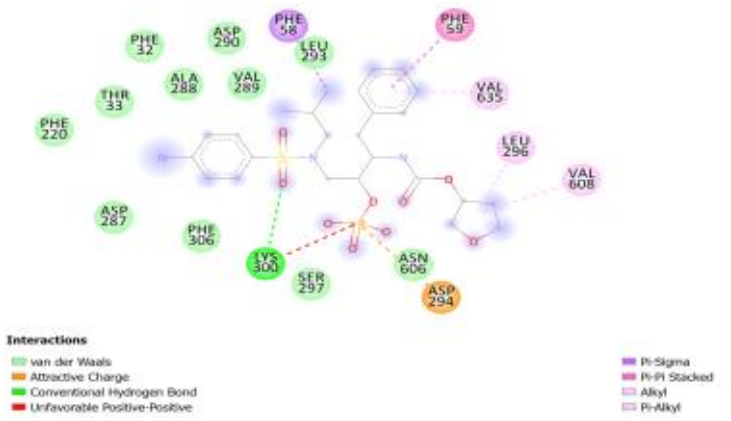 | Ala, Asn, 3Asp, 5Phe, 2Leu, Lys, Ser, 3Val | 1AC, 2Alk, 1HB, 1Pi-Alk, 1Pi-Pi-Stk, 1Pi-Sig, 1UPP, 11VW |
| 4 | Rintatolimod (C28H40N9O25P3), 135537060 | -7.6 | 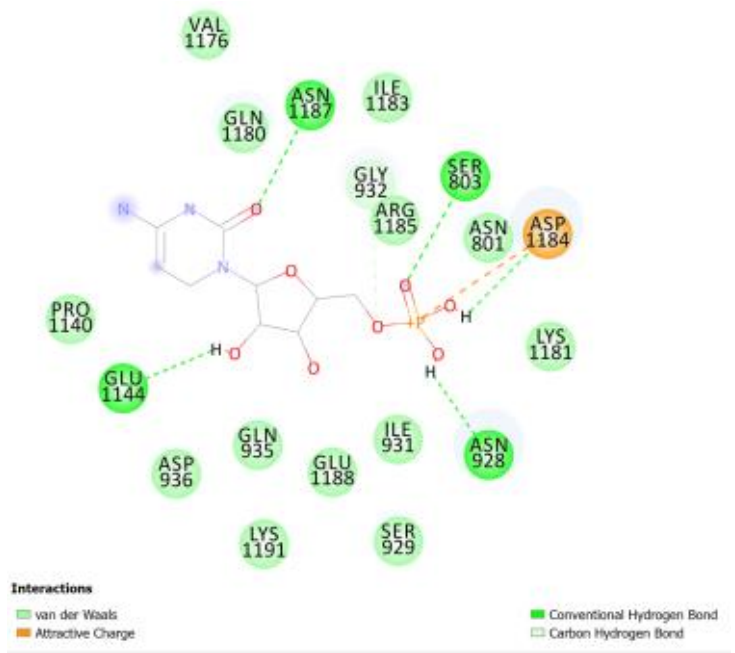 | Arg, 3Asn, 2Asp, 2Gln, 2Glu, Gly, 2Ile, 2Lys, Pro, 2Ser, 1Val | 1AC, 1CHB, 5HB, 14VW |
| 5 | Loviride (C17H16Cl2N2O2), 3963 | -7.3 | | Ala, Asn, 3Asp, Leu, | 2Alk, 1CHB, 3HB, 1Pi-Pi- |
| | | | 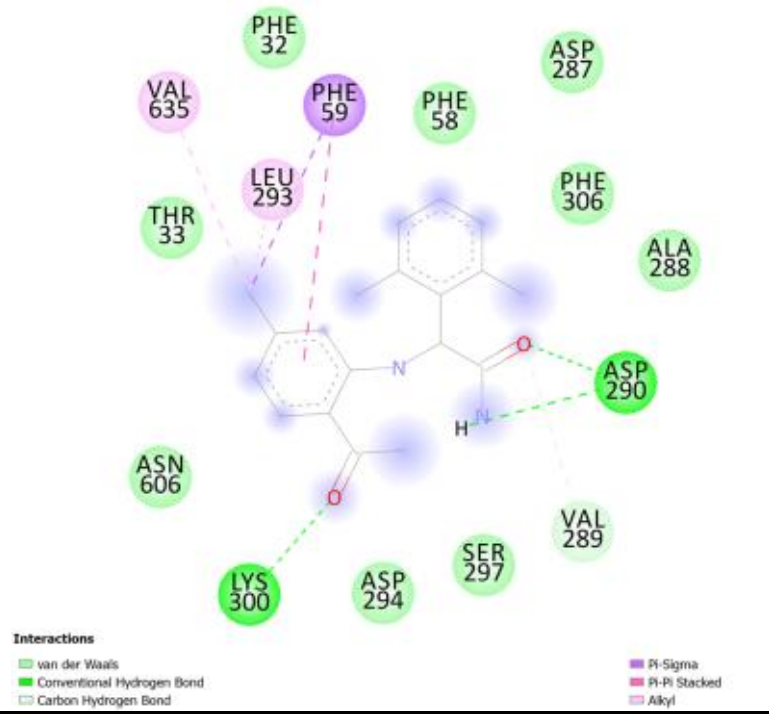 | Lys, 4Phe, Ser, Thr, 2Val | Stk, 1Pi-Sig, 9VW |
| 6 | Nevirapine (C15H14N4O), 4463 | -7.3 | 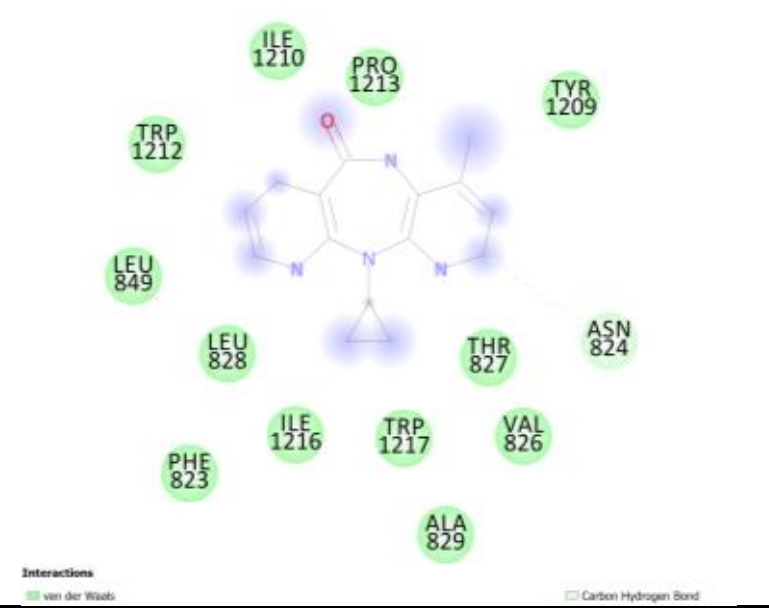 | Ala, Asn, 2Ile, 2Leu, Phe, Pro, Thr, 2Trp, Tyr, Val | 1CHB,12VW |
| 7 | Nitazoxanide (C12H9N3O5S), 41684 | -7.1 | 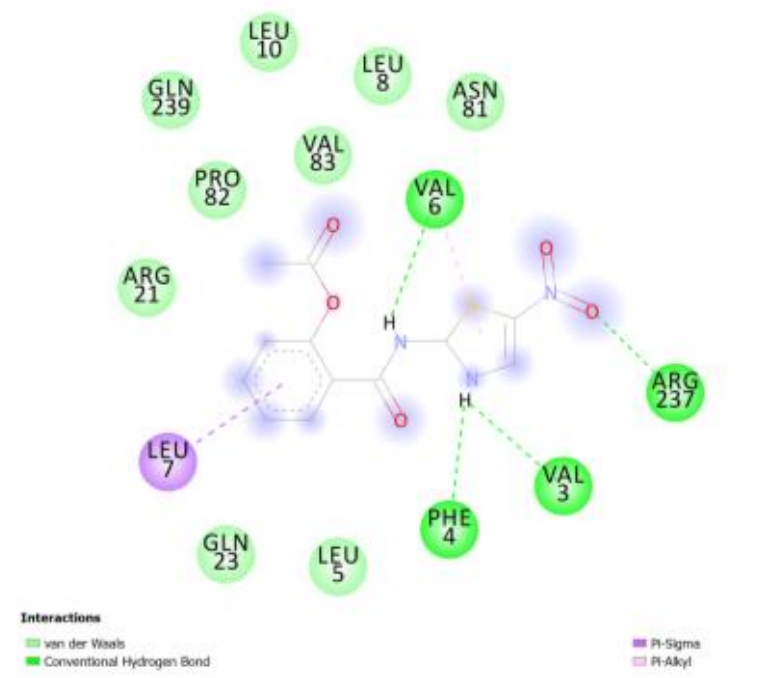 | 2Arg, Asn, 2Gln, Phe, Pro, 4Leu, 3Val | 4HB,12Pi-Alk, 1Pi-Sig, 9VW |
| 8 | Imiquimod (C14H16N4), 57469 | -6.8 | 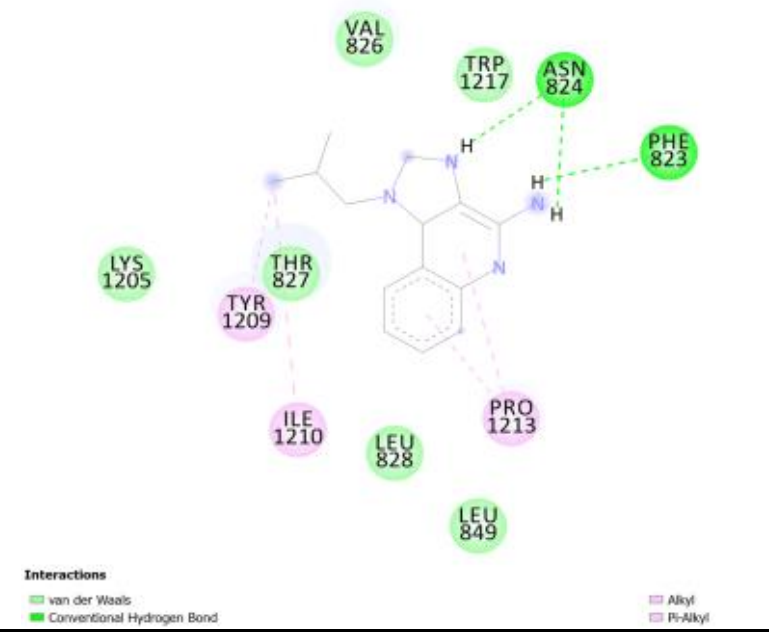 | Asn, Ile, 2Leu, Lys, Phe, Pro, Thr, Trp, Tyr, Val | 2Alk, 2HB, 1Pi-Alk, 5VW |
| 9 | Inosine (C10H12N4O5), 135398641 | -6.5 | | Arg, Asp, 5Leu, 2Phe, Pro, 3Val | 2HB, 12VW |
| | | | 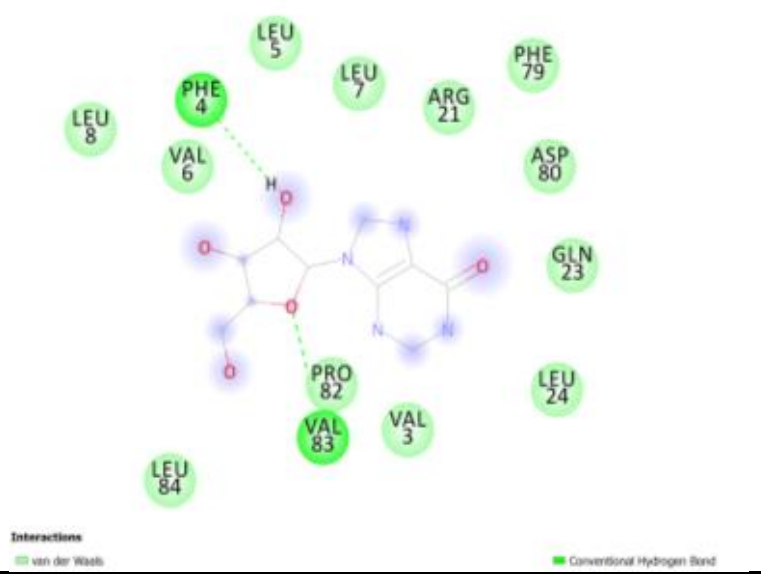 | | |
| 10 | Cobicistat (C40H53N7O5S2), 25151504 | -6.37 | 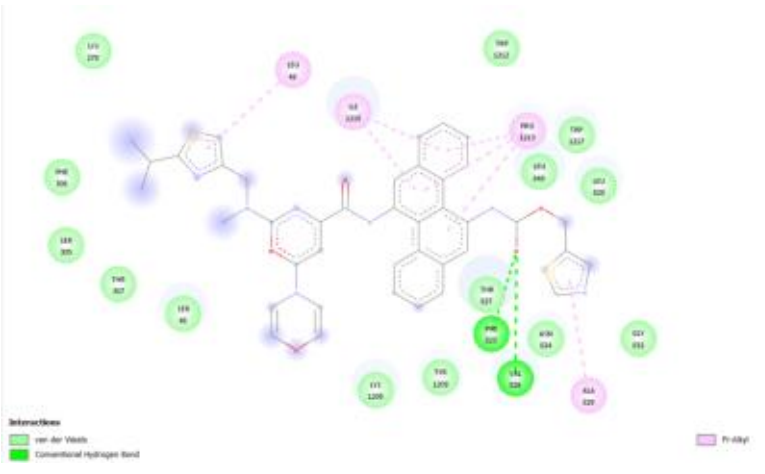 | 1Ala, 1Asn, 1Gly, 1Ile, 3Leu, 2Lys, 2Phe, 1Pro, 2Ser, 2Thr, 2Trp, 1Tyr, 1Val | 2HB, 7Pi-Alk, 14VW |

**Table S2:** List of the ten top hits of antibiotics drugs against SARS-CoV-2 Spike protein by order of lowest to highest binding energy value. Vancomycin seems to exhibit the lowest binding energy (-10.2 kcal/Mol) when docked with Spike protein, whereas Levofloxacin exhibits the highest binding energy (-5.11 kcal/Mol).

| | Antibiotics | Binding energy (kcal/Mol) | 2D interaction | Interface amino acids | Type of interactions |
|---|---|---|---|---|---|
| 1 | Vancomycin ($C_{66}H_{75}Cl_2N_9O_{24}$), 14969 | -10.2 | 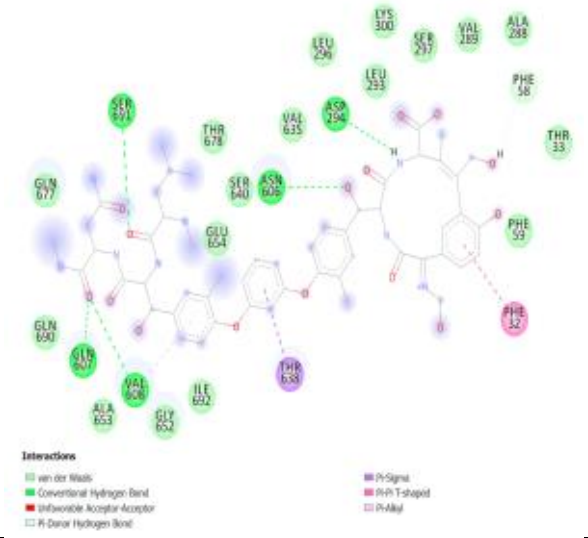 | 2Ala, Asn, Asp, 3Gln, Glu, Gly, Ile, 2Leu, Lys, 3Phe, 3Ser, 3Thr, 2Val | 5HB, 1Pi-Alk, 1Pi-DHB, 1Pi-Pi-T, 1Pi-Sig, 17VW |
| 2 | Gliclazide ($C_{15}H_{21}N_3O_3S$), 3475 | -8.6 | 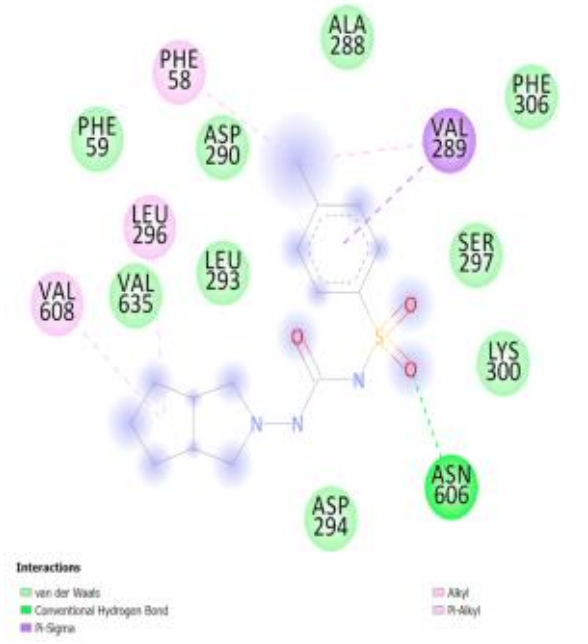  | Ala, Asn, 2Asp, 2Leu, Lys, 3Phe, Ser, 2Val | 3Alk, 1HB, 1Pi-Alk, 1Pi-Sig, 9VW |
| 3 | Azithromycin ($C_{38}H_{72}N_2O_{12}$), 447043 | -7.9 | 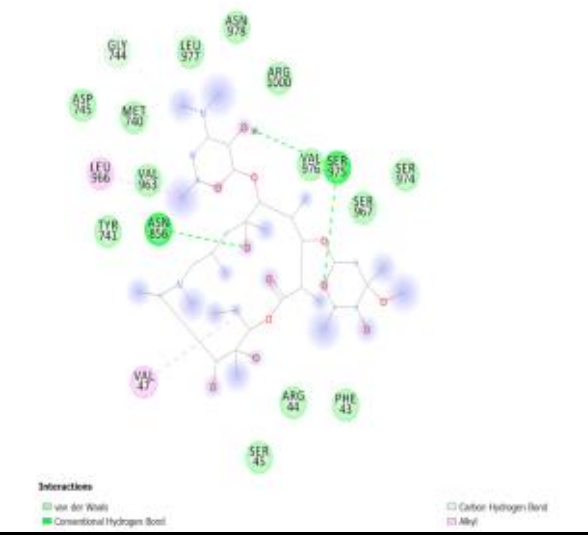 | 2Arg, 2Asn, Asp, Gly, 2Leu, Met, Phe, 4Ser, Tyr, 3Val | 2Alk, 1CHB, 3HB, 13VW |
| 4 | Sulfamethoxazole ($C_{10}H_{11}N_3O_3S$), 5329 | -6.6 | 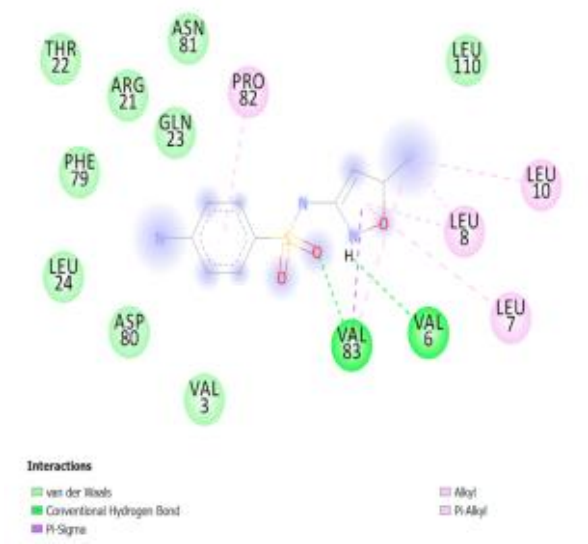  | Arg, Asn, Asp, Gln, 5Leu, Phe, Pro, Thr, 3Val | 3Alk, 2HB, 3Pi-Alk, 1Pi-Sig, 9VW |
| 5 | Meropenem ($C_{17}H_{25}N_3O_5S$), 441130 | -6.4 | 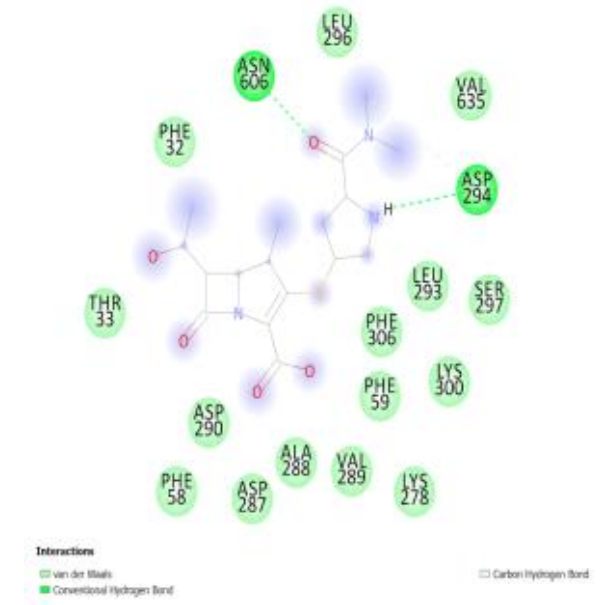 | Ala, Asn, 3Asp, 2Leu, 2Lys, 4Phe, Ser, Thr, 2Val | 1CHB, 2HB, 15VW |
| 6 | Tenofovir Disoproxil ($C_{19}H_{30}N_5O_{10}P$), 5481350 | -6.3 | 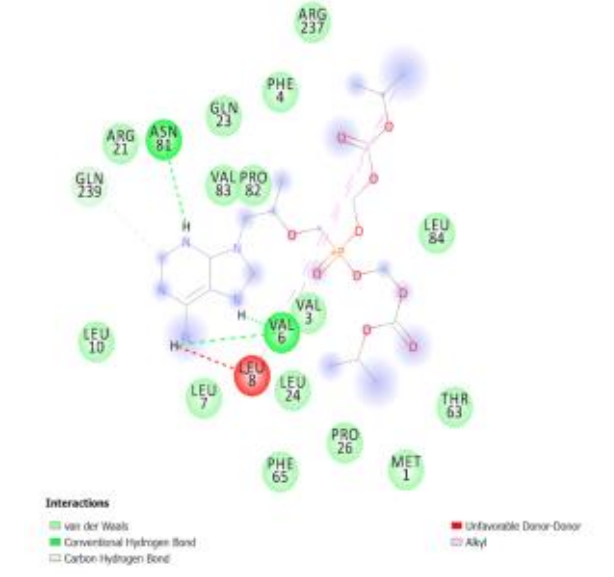 | 2Arg, Asn, 2Gln, 5Leu, Met, 2Phe, 2Pro, Thr, 3Val | 2Alk, 1CHB, 2HB, 1UDD, 16VW |
| 7 | Trimethoprim ($C_{14}H_{18}N_4O_3$), 5578 | -6.1 | 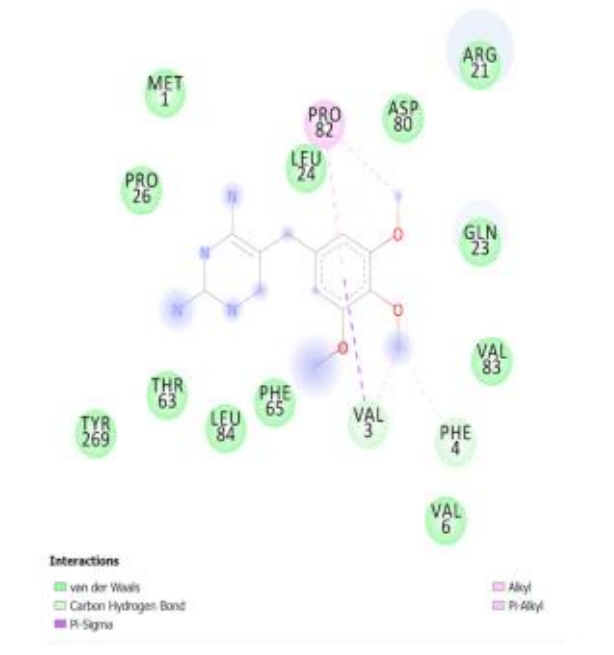 | Arg, Asp, Gln, 2Leu, Met, 2Phe, 2Pro, Thr, Tyr, 3Val | 2Alk, 2CHB, 1Pi-Alk, 1Pi-Sig, 12VW |
| 8 | Ciprofloxacin ($C_{17}H_{18}FN_3O_3$), 2764 | -5.58 | 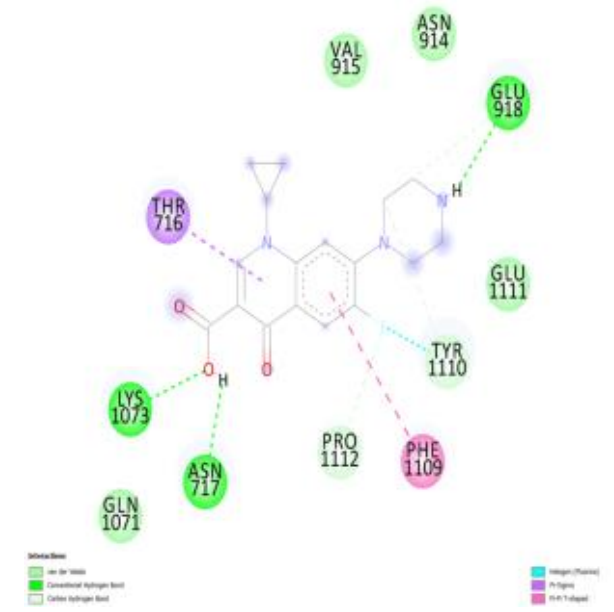 | 2Asn, 1Gln, 2Glu, 1Lys, 1Phe, 1Pro, 1Thr, 1Tyr, 1Val | 3CHB, 1Hal, 3HB, 1Pi-Pi-T, 1Pi-Sig, 4VW |
| 9 | Gentamicin ($C_{21}H_{43}N_5O_7$), 3467 | -5.4 | 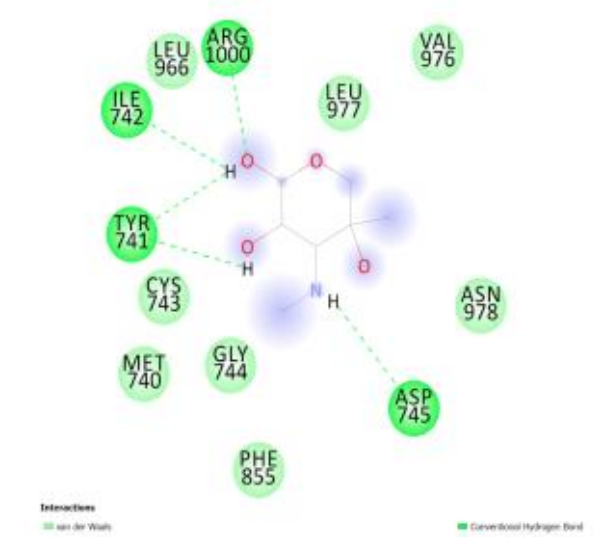 | Arg, Asn, Asp, Cys, Gly, Ile, 2Leu, Met, Phe,Tyr, Val | 5HB, 8VW |
| 10 | Levofloxacin ($C_{18}H_{20}FN_3O_4$), 149096 | -5.11 | 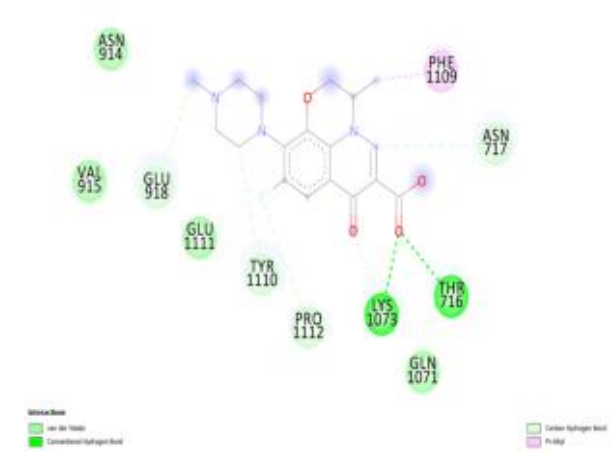 | 2Asn, 1Gn, 2Glu, 1Lys, 1Phe, 1Pro, 1Thr, 1Tyr, 1Val | 5CHB, 2HB, 1Pi-Alk, 4VW |

**Table S3:** List of the top hits antiparasitic drugs, flavonoids, and Vitamins against SARS-CoV-2 Spike protein by order of lowest to highest binding energy value. Ivermectin B1a seems to exhibit the lowest binding energy (-9.16 kcal/Mol) when docked with Spike protein, whereas Hydroxychloroquine exhibits the highest binding energy (-2.85 kcal/Mol). In terms of flavonoids Tetramethoxyflavone exhibit, the lowest binding energy (-4.91 kcal/Mol) and Gallocatechin exhibits the highest binding energy (-2.88). Moreover, Vitamin D showed the lowest binding energy (-5.52 kcal/Mol) and Vitamin C the lowest binding energy (-2.95 kcal/Mol).

| | **Antiparasitic** | **Binding energy (kcal/Mol)** | **2D interaction** | **Interface amino acids** | **Type of interactions** |
|---|---|---|---|---|---|
| 1 | Ivermectin B1a ($C_{48}H_{74}O_{14}$), 6321424 | -9.16 | 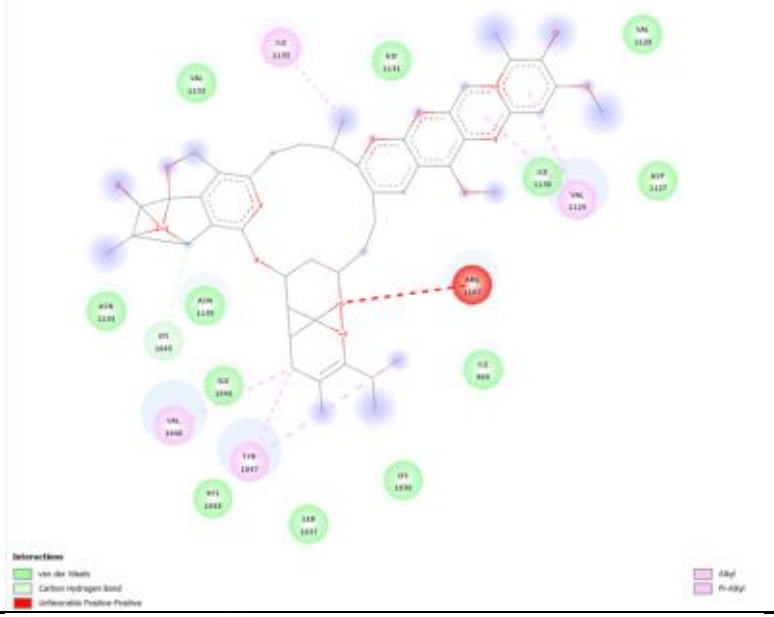 | 1Arg, 2Asn, 1Asp, 2Gly, 1His, 3Ile, 2Lys, 1Ser, 1Tyr, 4Val | 4Alk, 1CHB, 2Pi-Alk, 1UPP, 12VW |
| 2 | Ivermectin B1b ($C_{47}H_{72}O_{14}$), 6321425 | -8.86 | 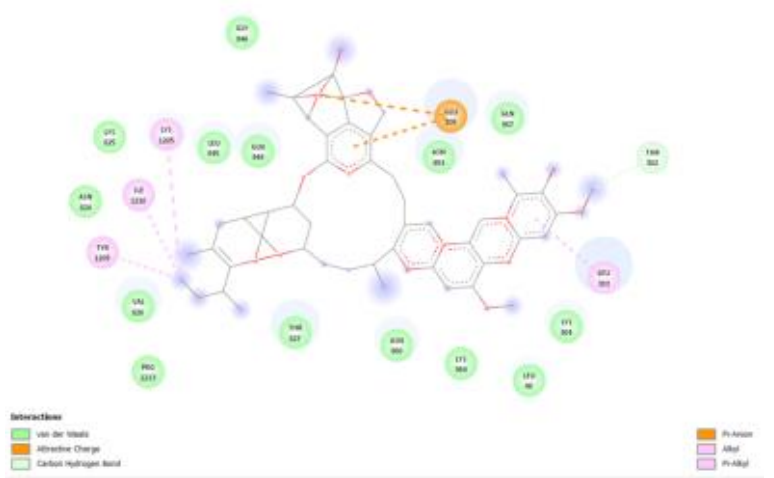 | 3Asn, 2Gln, 1Glu, 1Gly, 1Ile, 3Leu, 4Lys, 1Pro, 2Thr, 1Tyr, 1Val | 2AC, 3Alk, 1CHB, 1Pi-Alk, 14VW |
| 3 | Hydroxychloroquine ($C_{18}H_{26}ClN_3O$), 3652 | -2.85 | 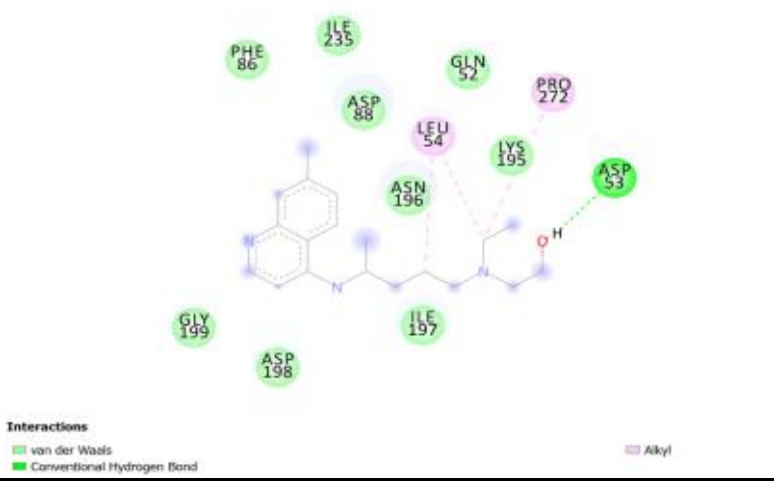 | Asn, 3Asp, Gln, Gly, 2Ile, Leu, Lys, Phe, Pro | 2Alk, 1HB, 9VW |
| | | | | | |
| | **Flavonoids** | **Binding energy (kcal/Mol)** | **2D interaction** | **Interface amino acids** | **Type of interactions** |
| 1 | Tetramethoxyflavone ($C_{19}H_{18}O_6$), 471721 | -4.91 | 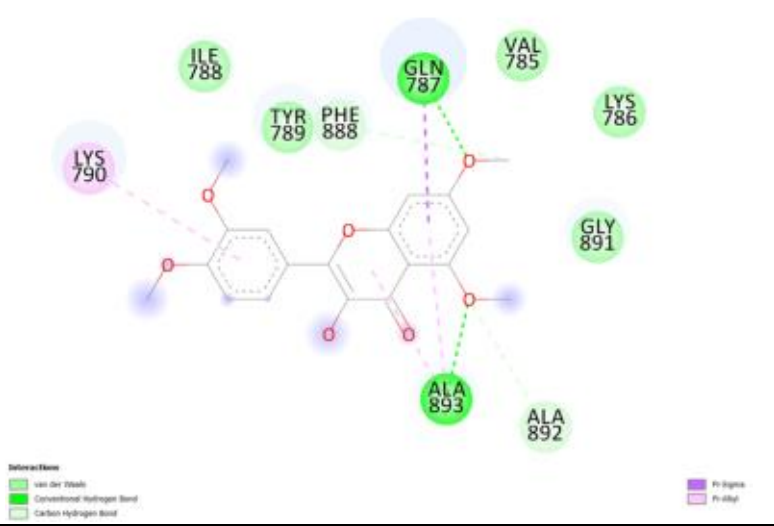 | 2Ala, 1Gln, 1Gly, 1Ile, 2Lys, 1Phe, 1Tyr, 1Val | 2CHB, 2HB, 3Pi-Alk, 1Pi-Sig, 5VW |
| 2 | Herbacetin (C15H10O7), 5280544 | -4.74 | 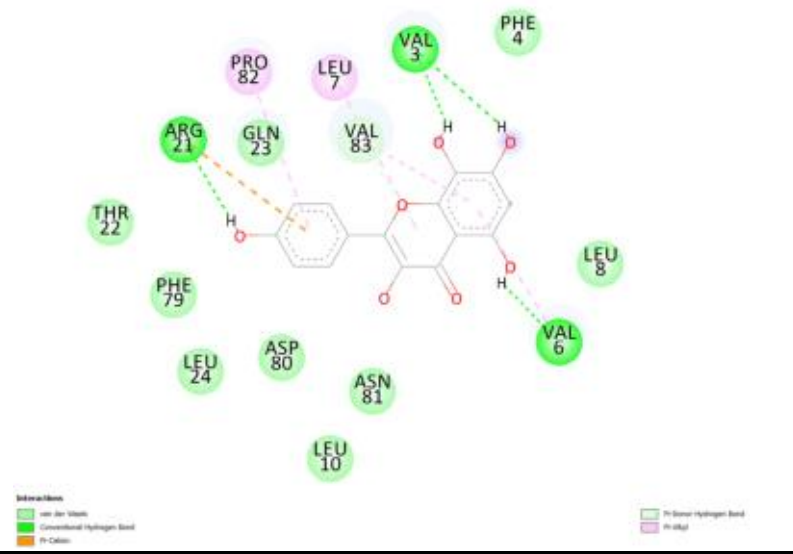 | 1Arg, 1Asn, 1Asp, 1Gln, 4Leu, 2Phe, 1Pro, 1Thr, 3Val | 4HB, 4Pi-Alk, 1Pi-Cat, 1Pi-DHB, 9VW |
| 3 | Gallocatechin (C15H14O7), 65084 | -2.88 | 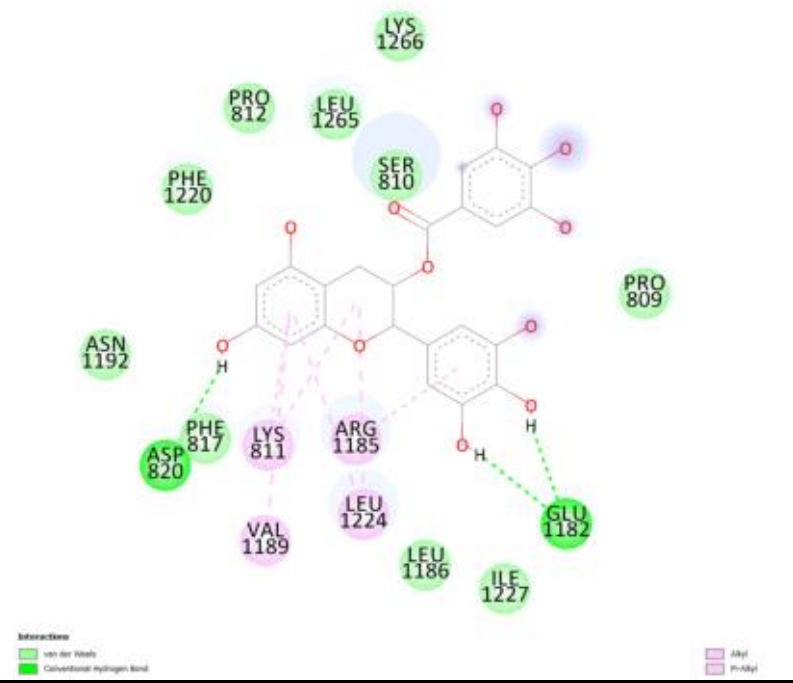 | 1Arg, 1Asn, 1Asp, 1Glu, 1Ile, 3Leu, 2Lys, 1Ser, 2Phe, 2Pro, 1Val | 2Alk, 3HB, 4Pi-Alk, 10VW |
| | | | | | |
| | **Vitamins** | **Binding energy (kcal/Mol)** | **2D interaction** | **Interface amino acids** | **Type of interactions** |
| 1 | Vitamin D (C27H44O), 5280795 | -5.52 | 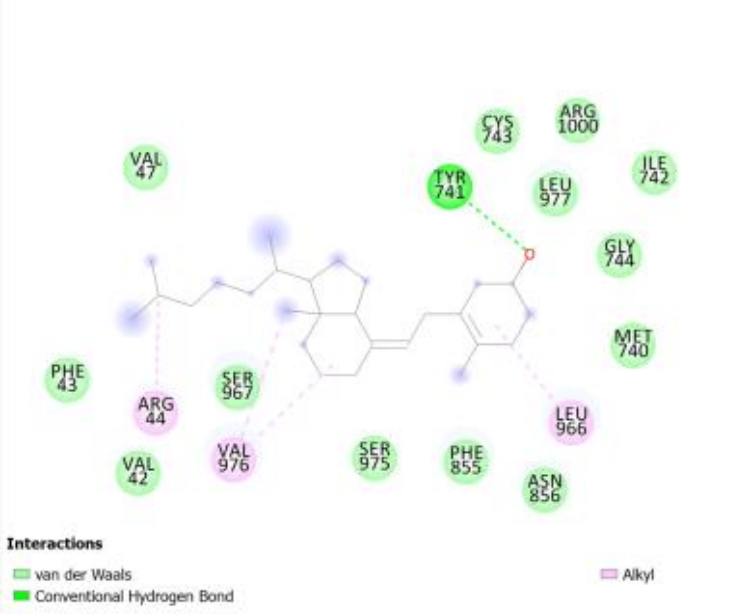 | 2Arg, Asp, Cys, Gly, Ile, 2Leu, Met, 2Phe, 2Ser, Tyr, 3Val | 3Alk, 1HB, 12VW |
| 2 | Vitamin C (C6H8O6), 54670067 | -2.95 | 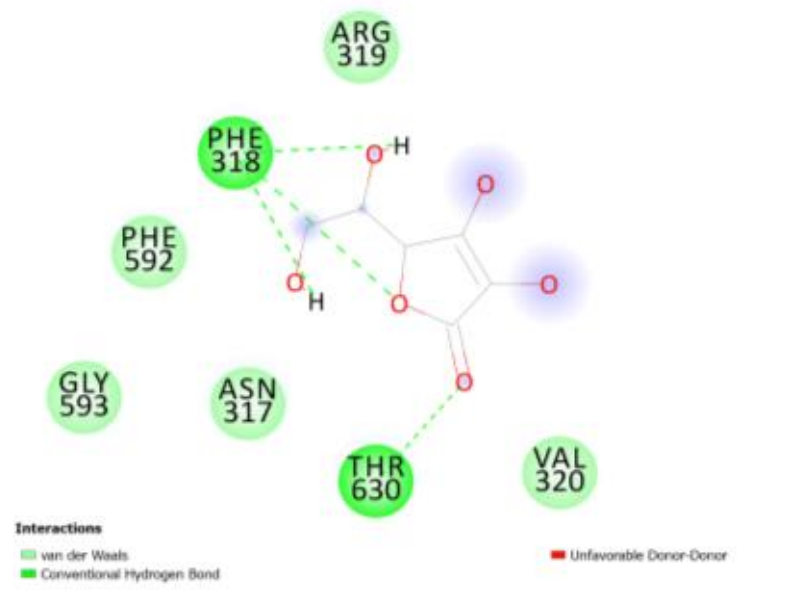 | Arg, Asn, Gly, 2Phe, Thr, Val | 4HB, 5VW |

**Table S4**: A comparative analysis between SARS-CoV-2 RdRP binding affinities and spike protein binding affinities in antivirals which primarily target viral RdRP. These comparisons include binding energy, the specific amino acids that are interacting with the drug, and the types of bonds formed during this interaction.

| | **Name and PubChem ID** | **Bound molecule** | **Binding Energy (kcal)** | **2D Interaction Plot** | **Interface Amino Acid** | **Type of Interactions** |
|---|---|---|---|---|---|---|
| 1 | Beclabuvir (C36H45N5O5S), 49773361 | RdRp | -5.63 | 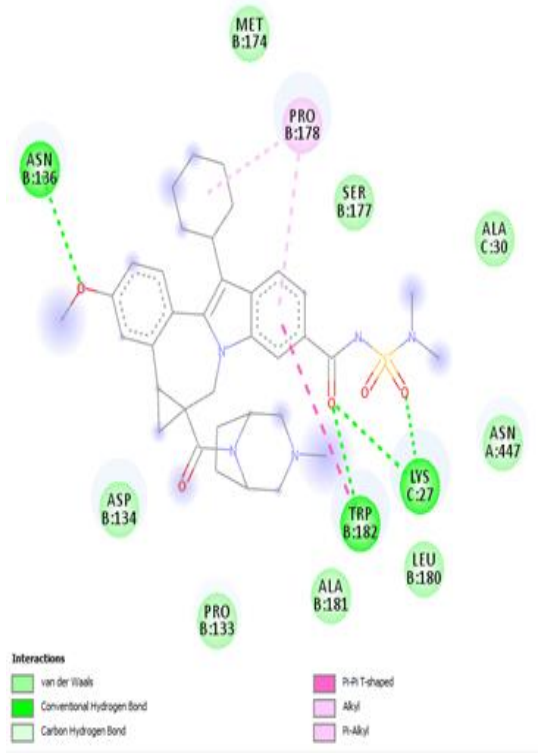  | 2Ala, 2Asn, Asp, Leu, Lys, Met, 2Pro, Ser, Trp | 4HB, 2Pi-Alk, 1Pi-Pi-T, 8VW |
| | | Spike | -6.64 | 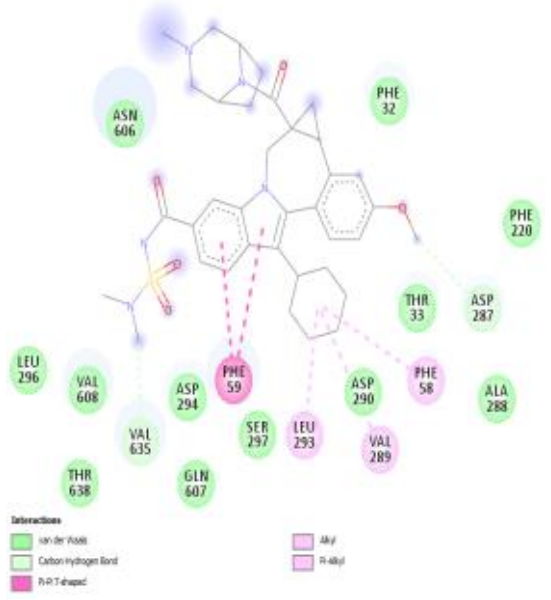  | Ala, Asn, 3Asp, Gln, 2Leu, 4Phe, Ser, 2Thr, 3Val | 2Alk, 2CHB, 1Pi-Alk, 2Pi-Pi-T; 12VW |
| 2 | Galidesivir (C11H15N5O3), 10445549 | RdRp | -4.38 | 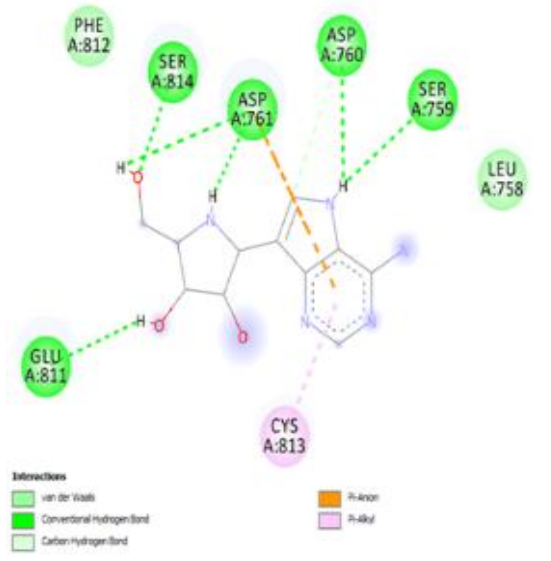  | 2Asp, Cys, Glu, Leu, Phe, 2Ser | 1CHB, 6HB, 1Pi-Alk, 1Pi-Ani, 2VW |
| | | Spike | -3.11 | 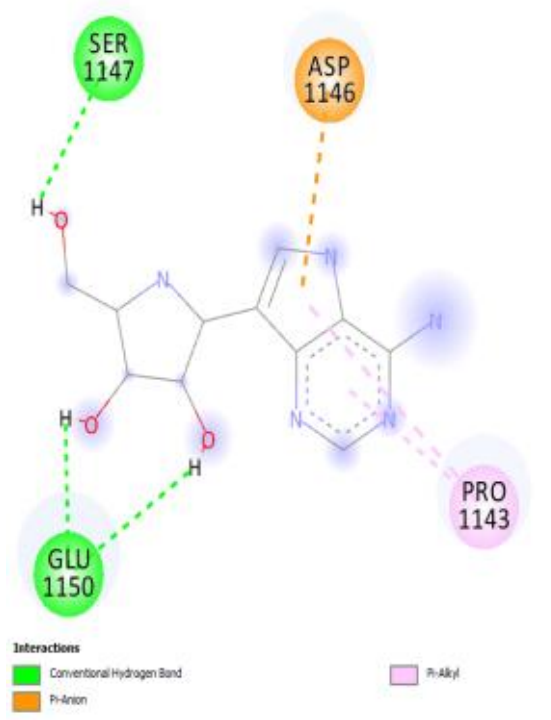  | Asp, Glu, Pro, Ser | 3HB, 2Pi-Alk, 1Pi-An |
| 3 | Ribavirin (C8H12N4O5), 37542 | RdRp | -4.2 | 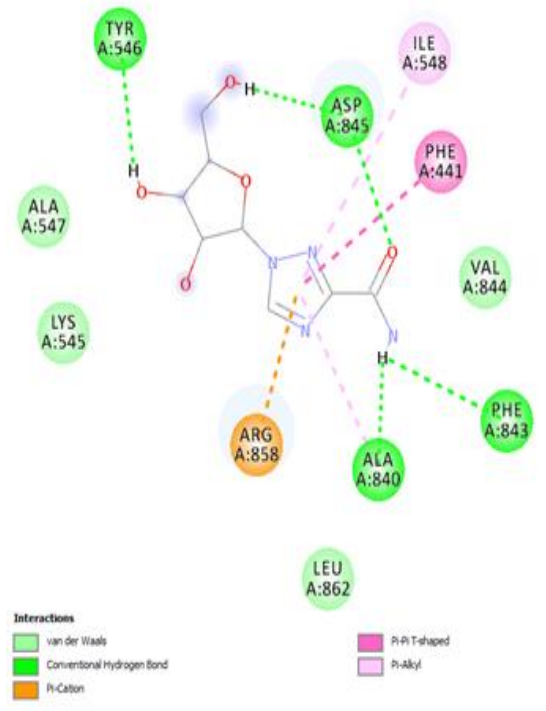 | 2Ala, Arg, Asp, Ile, Leu, Lys, 2Phe, Val | 5HB, 2Pi-Alk, 1Pi-Cat, 1Pi-Pi, 4VW |
| | | Spike | -3.29 | 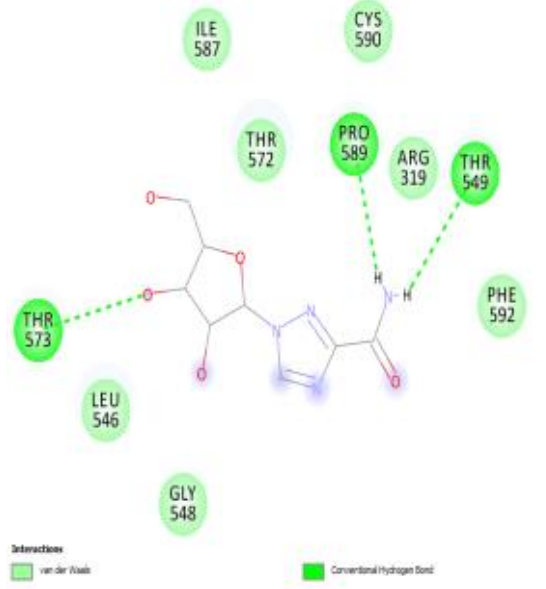 | Arg, Cys, Gly, Ile, Leu, Phe, Pro, 3Thr | 3HB, 7VW |
| 4 | Favipiravir (C5H4FN3O2), 492405 | RdRp | -3.09 | 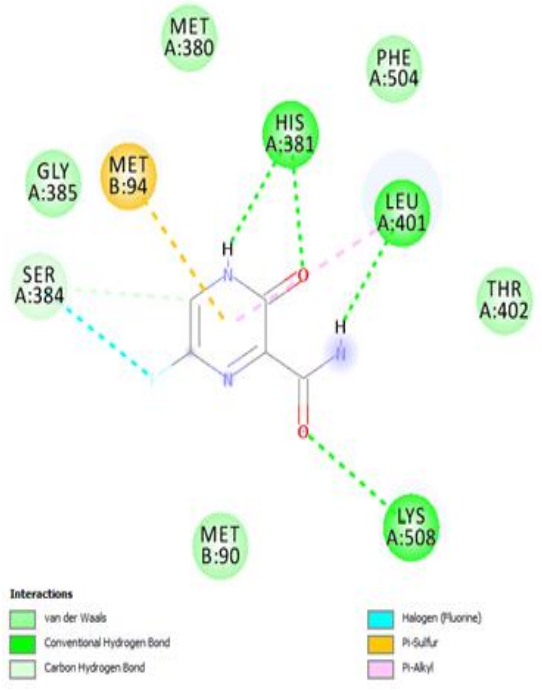 | Gly, His, Leu, Lys, 3Met, Phe, Ser, Thr | 1CHB, 1Hal, 4HB, 1Pi-Alk, 1Pi-Sul, 5VW |
| | | Spike | -2.94 | 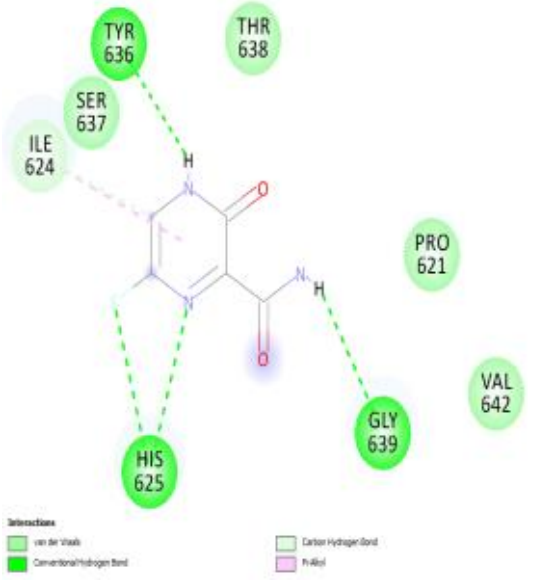 | Gly, His, Ile, Pro, Ser, Thr, Tyr, Val | 1CHB, 4HB, 1Pi-Alk, 4VW |
| 5 | Sofosbuvir ($C_{22}H_{29}FN_3O_9P$), 45375808 | RdRp | -2.95 | 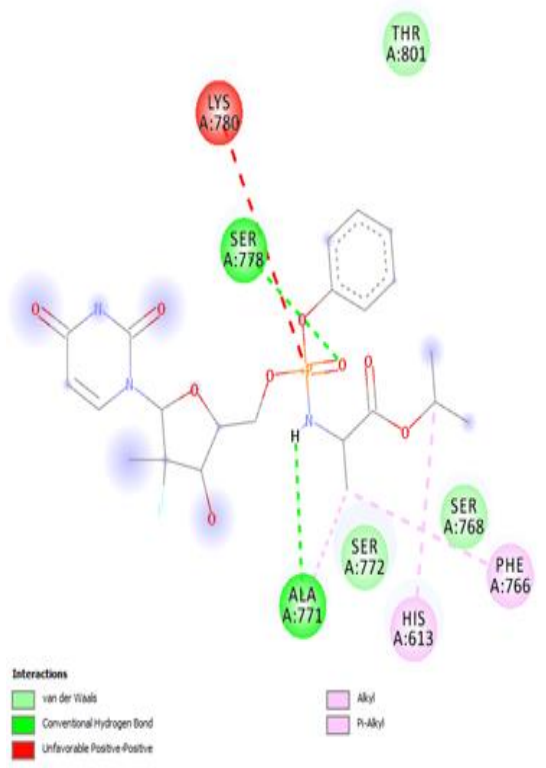 | Ala, His, Lys, Phe, 3Ser, Thr | 1Alk, 2HB, 2Pi-Alk, 3VW, 1UPP |
| | | Spike | -3.7 | 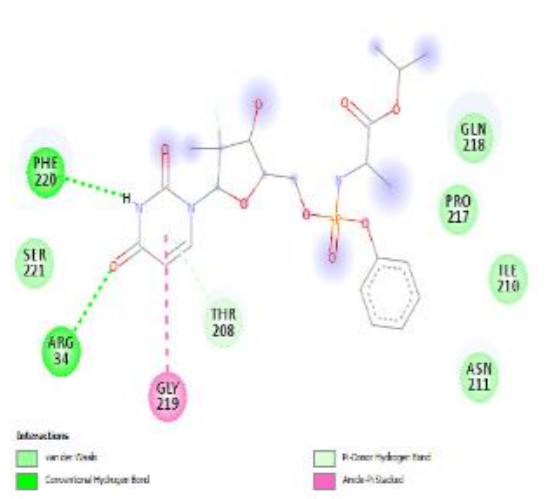 | Arg, Asn, Gln, Gly, Ile, Phe, Pro, Ser, Thr | 1APS, 2HB, 1Pi-DHB, 5VW |
| 6 | Tenofovir ($C_9H_{14}N_5O_4P$), 464205 | RdRp | -2.34 | 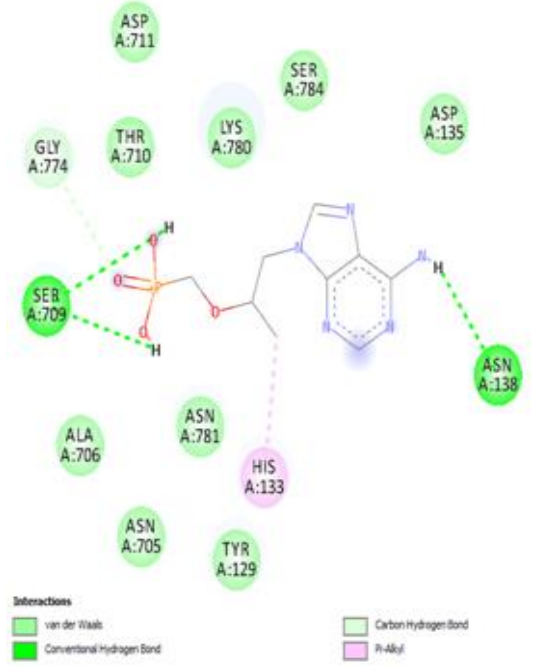 | Ala, 3Asn, 2Asp, Gly, His, Lys, 2Ser | 1CHB, 3HB, 1Pi-Alk, 9VW |
| | | Spike | -2.75 | 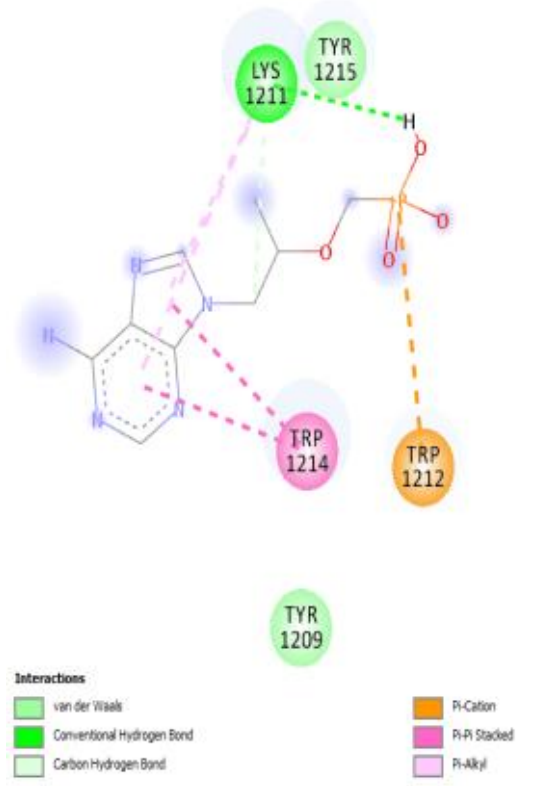 | Lys, 2Trp, 2Tyr | 1CHB, 1HB, 2Pi-Alk, 1Pi-Cat, 2Pi-Pi-Stk, 2VW |
| 7 | Remdesivir ($C_{27}H_{35}N_6O_8P$), 121304016 | RdRp | -1.28 | 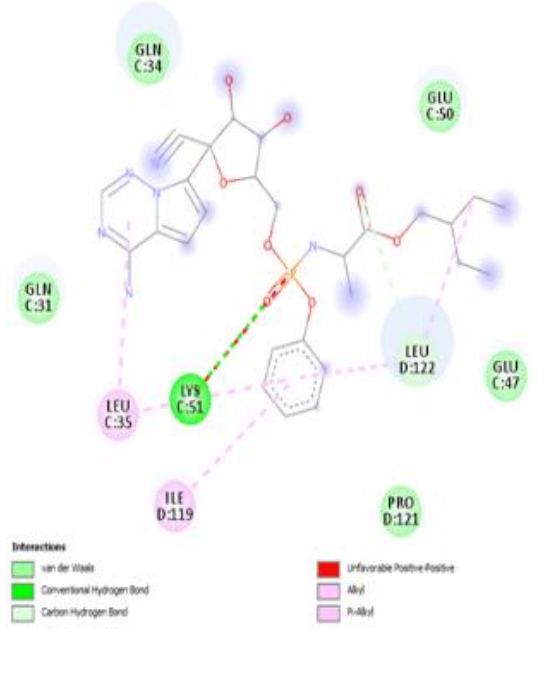 | 2Gln, 2Glu, Ile, 2Leu, Lys, Pro | 1Alk, 1CHB, 1HB, 4Pi-Alk, 5VW, 1UPP |
| | | Spike | -2.65 | 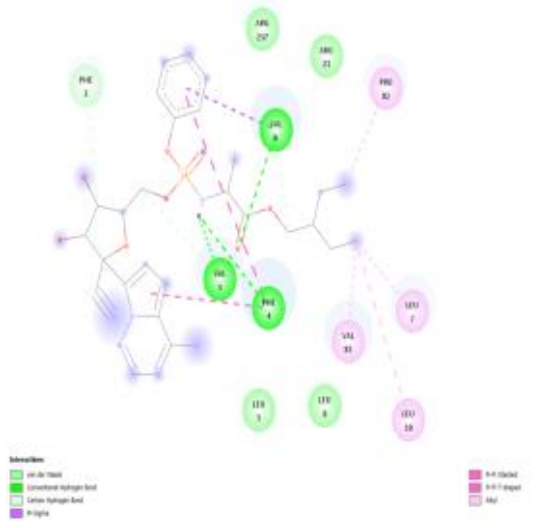 | 2Arg, 4Leu, 2Phe, 1Pro, 3Val | 4Alk, 1CHB, 3HB, 1Pi-Pi-T, 1Pi-Pi-Stk, 1Pi-Sig, 4VW |